\documentclass[12pt]{article}

\usepackage{setspace}
\usepackage{amsmath,amsfonts}
\usepackage{theorem}

\usepackage{epsfig}

\usepackage{xcolor}

\usepackage{lscape}

\usepackage{multirow}

\usepackage{rotating}

\usepackage{tikz}

\usepackage{makecell}
\newcolumntype{x}[1]{>{\centering\arraybackslash}p{#1}}
\usepackage[hyphens]{url}
\usepackage[breaklinks]{hyperref}
\hypersetup{
    colorlinks=true,       
    linkcolor=red,          
    citecolor=blue,        
    filecolor=magenta,      
    urlcolor=cyan           
}

\usepackage[sort&compress]{natbib}
\bibpunct{(}{)}{,}{a}{,}{,}

\theoremstyle{plain} { \theorembodyfont{\rmfamily}

}
\newtheorem{proposition}{Proposition}[section]
\newtheorem{lemma}{Lemma}[section]

\newtheorem{theorem}{Theorem}[section]

\def\halmos{\mbox{\quad$\square$}}
\def\qed{\quad Q.E.D.}

\newenvironment{pfof}[1]{\vspace{1ex}\noindent{\bf Proof of #1}\hspace{0.5em}}
    {\qed\vspace{1ex}}

\newcommand{\cF}{{\cal F}}

\newcommand{\bN}{\mathbb{N}}

\newcommand{\bZ}{\mathbb{Z}}
\newcommand{\prob}{\mathbb{P}}
\newcommand{\expect}{\mathbb{E}}

\begin{document}
\title{Optimal exit strategies of CPT gamblers in unfair  gambles \thanks{
The first author is funded by National Natural Science of China (Grant No. 12271462) and the Fundamental Research Funds for the Central Universities of Shanghai University of Finance and Economics, and the second author by the Nie Center for Intelligent Wealth Management at Columbia.  The authors are grateful to Jan Ob{\l}{\'o}j for an early discussion on the work.}
}

\author{Sang Hu\thanks{Institute of Big Data Research, School of Statistics and Data Science, Shanghai University of Finance and Economics, Shanghai, China 200433.
Email: \texttt{husang@sufe.edu.cn}.}
\and Xun Yu Zhou\thanks{Department of Industrial Engineering and Operations Research, Columbia University, New York, New York 10027.
Email: \texttt{xz2574@columbia.edu}.}
}

\maketitle

\begin{abstract}
In this paper we study optimal exit strategies of gamblers with cumulative prospect theory (CPT) preferences in games where the expected payoff is strictly negative at each play, and  formulate the problem as optimal stopping on asymmetric random walks.  Applying a geometric transformation of the underlying cumulative gain/loss process, engaging randomized strategies and changing the decision variable from stopping times to probability distribution of the accumulated gain or loss at exit time, we solve the problem via the Skorokhod embedding.
Drastically different from the fair gamble problem studied by \cite{HeEtal2019:StoppingStrategies}, we show that the unfair problem in the infinite time horizon has finite values for a wide range of CPT parameter specifications.
We then present the analytical solutions  in the case of piece-wise power utility and power probability distortion functions.
Compared to the strategies used in fair gambling, the CPT gamblers in unfair gambles are less loss-tolerant and choose not to gamble at all when the games are sufficiently unfavorable. 

\bigskip

\end{abstract}

\section{Introduction}\label{se:Introduction}

Casino gambling has long been a focus of economic research, partly because gambling, as a risk-seeking endeavor, contradicts the risk-averse rational assumption inherent in the classical expected utility theory (EUT). Under EUT, people's utility functions are concave. Therefore, Jensen's inequality dictates that,  the expected payoff of a {\it fair} game being zero, the utility of not participating in gambling is greater than the utility of participating. In reality, considering factors such as casino labor costs and equipment operating costs and that casinos aim to make profits, gamblers always  face unfair/disadvantageous games whose expected payoffs are negative yet still participate, further contradicting EUT. Some researchers have attempted to explain this using non-concave utility functions or additional gambling utility models, but these approaches are overly specific.
\cite{Barberis2012:Casino} first proposes using the cumulative prospect theory (CPT; \citealp{TverskyKahneman1992:CPT})  from behavioral finance to study casino gambling and explains within this theoretical framework why people participate in games with zero expected payoff. Intuitively, the probability distortion factor in CPT amplifies the weight of low-probability events like ``winning big,''  allowing gamblers to obtain random payoffs with skewed probability distributions by choosing appropriate exit strategies. This is because random payoffs with skewed probability distributions correspond to higher or even positive CPT values. Thus, even if the expected return is zero, a gambler will participate in casino gambling seeking a positive CPT value.

However, due to probability distortion, the objective of the problem under CPT is nonlinear with respect to the probability distribution. This makes a multi-period gambling stopping problem unsolvable using traditional techniques such as dynamic programming and martingale methods. While exhaustive search can be used to obtain the optimal stopping strategy in a five-period symmetric random walk model in \cite{Barberis2012:Casino}, further extending the time period significantly increases the computational load, rendering exhaustive search inapplicable. To systematically solve the optimal stopping problem in gambling under CPT, \cite{HeEtal2019:StoppingStrategies} apply the Skorokhod embedding to transform the decision variable from stopping time to the probability distribution function of cumulative profit or loss at the stopping time. Assuming a simple symmetric random walk over an infinite time period, they derive the optimal stopping strategy under CPT. Subsequently, \cite{HuEtal2022:ACasinoGamblingModel} investigate stopping strategies under a finite time horizon.

Unlike \cite{Barberis2012:Casino}, which restricts attention to Markovian deterministic strategies, \cite{HeEtal2019:StoppingStrategies} allow for both path-dependent and randomized strategies. In other words, the decision of whether to stop at a given time depends not only on the current gain or loss position, but also on the entire past trajectory, and may even depend on an external random variable that is independent of the underlying random walk such as the outcome of a coin toss.
From a mathematical perspective, \cite{HeHuOblojZhou2017:Twoexplicitembedding} show that introducing randomized stopping times convexifies the set of attainable probability distributions, leading to necessary and sufficient conditions for the Skorokhod embedding to work  in a simple symmetric random walk.
Moreover, in practice, people do use randomization devices such as coin tosses to assist decision making. \cite{HeHuOblojZhou2016:Randomization} show that, under CPT, path-independent strategies are dominated by path-dependent strategies, while deterministic strategies are further dominated by randomized strategies. Therefore, randomized strategies are optimal when they can be executed.

\cite{Barberis2012:Casino} and the series of works of \cite{HeHuOblojZhou2016:Randomization,HeEtal2019:StoppingStrategies,
HeHuOblojZhou2017:Twoexplicitembedding,HuEtal2022:ACasinoGamblingModel}
 consider fair symmetric gambling model in which the gambler's expected payoff from gambling is zero. In reality, casinos need to make profits; hence casino gambling odds are always set to be unfavorable to gamblers. In other words, gamblers are expected to lose money.
This work will investigate the behavior of gamblers with CPT preferences in an asymmetric, unfair gambling model. The central question is whether
the gambler will enter such an unfair game and what the optimal exit time is if he does enter.

Our contribution is threefold. First and technically, the key Skorokhod embedding technique in the previous study on fair games no longer works directly for the unfair counterpart due to the loss of martingality.
To overcome the difficulty, we consider a geometric transformation on the underlying gain/loss process to obtain a new martingale and then prove Skorokhod embedding theorem by designing randomized Az\'ema--Yor stopping times.
Second, we discover that the optimal values of asymmetric gambling in infinite horizon are finite for a wide range of model specifications, in sharp contrast to the infinite objective value of symmetric gambling under the same specifications. Third and economically, the gamblers are less loss-tolerant and the optimal strategies are more conservative in asymmetric gambling than in symmetric ones under the same parameter settings. The gambler may still enter into an unfair game because he can construct certain stopping strategy such that the probability distribution of his payoff from gambling is positively-skewed and thus preferred due to probability distortion. However, the unfavorable gambling odds makes him less patient to stay in the game when his loss relative to the historical maximum wealth is large. Indeed, the longer the gambler stays in the casino, the more he is expected to lose.

The remainder of the paper is organized as follows. In Section \ref{se:Model}, we introduce the unfair asymmetric gambling model and formulate the corresponding optimal stopping problem. In Section \ref{se:Skorokhodembedding}, we provide necessary and sufficient conditions to prove the Skorokhod embedding theorem and illustrate randomized Az\'ema--Yor stopping times. In Section \ref{se:indp}, we reformulate the original optimal stopping problem into an infinite-dimensional programming and show that the problem has finite optimal value for broad model specifications. In Section \ref{se:powerutilityanddistortion}, we solve the problem analytically with piece-wise power utility and power probability distortion functions. Section \ref{se:UnfairGamesNumericalExample} provides numerical examples and a comparison to the strategies in symmetric gambling. Section \ref{se:UnfairGamesConcludingRemarks} concludes.

\section{Model Formulation}\label{se:Model}

\subsection{Setting}
Assume that a gambler enters a casino for free.
At time 0 the gambler is faced with a bet: with probability $p$ to win one dollar or probability $1-p$ to lose one dollar at time 1.
Here $0 < p < 1/2$ so that the probability of winning one dollar is strictly smaller than that of losing one dollar in a bet. As a result, the expected payoff from the bet is negative to the gambler.
This is consistent with the real situation in which gamblers are almost always faced with unfavorable odds.
The gambler can choose to accept or reject this bet. If he accepts it, the bet is played out at time 1 and he is faced with the next bet; if he rejects it, he quits and leaves the casino.
Suppose the same bet is offered to the gambler sequentially at time $t = 0,1,...$ if he does not quit.
If the gambler rejects the bet at some time $t$, then he stops at $t$. 

\subsection{Preference}\label{subse:Preference}

The gambler has the risk preference of CPT of \cite{TverskyKahneman1992:CPT}.
There are four key features of CPT, making it very different from the classical EUT.
First, a reference point is used to measure the relative gain or loss of the total payoff.
Second, the utility function is applied to the gain or loss rather than to the total payoff and the marginal utility is decreasing.
Third, the disutility from losses is greater than the utility from gains with same magnitude, a property called loss-aversion.
A typical utility function is therefore S-shaped and has a kink at the reference point.
Forth, a probability distortion/weighting function is present in the preference.
The probability weighting function is typically inverse S-shaped so that events with small probabilities are usually overweighted, while events with moderate probabilities are underweighted.
This last feature makes the CPT preference value no longer an objective expectation over all the possible scenarios, but nonlinear in the probability distribution of the random payoff, in a sharp contrast to the EUT preference.

Specifically, with CPT preference a random payoff $X$ is evaluated as
\begin{align}\label{eq:CPTfunctional}
{\cal V}(X) = \int_{B}^{+\infty} u(x-B) d [1 - w_+(1 - F_X(x))] + \int_{-\infty}^{B} u(x-B) d [ w_-(F_X(x)) ],
\end{align}
where $B$ denotes the reference point, $u(\cdot)$ is the utility function with $u(0) = 0$, and $w_\pm(\cdot) : [0,1] \mapsto [0,1]$ are the probability weighting functions applied in the gain region and the loss region, respectively.

\subsection{Optimal stopping}\label{subse:Optimalstopping}
Suppose the gambler has initial wealth $X_0$. Denote by $X_t$ his total wealth at time $t$. 
He needs to decide whether he would like to take a bet and, if he does, what the best time is to exit.
Let $\tau : \Omega \mapsto \{0,1,2,...\}  \cup \{\infty\}$, adapted to some filtration (which is to be defined later).
The initial wealth is naturally a reference point to define gains and losses, which we assume to be the case in this study.
Let $S_t : = X_t - X_0$ be his cumulative gain or loss up to time $t$.
The gambler's decision problem is then formulated below as an optimal stopping problem to maximize the CPT value of his gain/loss at exit time over all the possible stopping strategies:
\begin{align}\label{optprob}
\sup_{\tau \in \cal T} {\cal V}(S_\tau), \tag{P}
\end{align}
where $\cal T$ denotes the set of admissible stopping strategies to be defined momentarily.

In view of  the CPT functional \eqref{eq:CPTfunctional}, we write the objective ${\cal V}(S_\tau)$ in our optimal stopping problem \eqref{optprob} to be
\begin{equation}\label{objtfunc}
\begin{aligned}
V(S_\tau) &: = \sum_{n=1}^\infty u_+(n)\Big(w_+ \big(\prob(S_\tau \geq n) \big) - w_+ \big(\prob(S_\tau \geq n+1) \big)\Big) \\
    & \quad - \sum_{n=1}^\infty u_-(n)\Big(w_- \big(\prob(S_\tau \leq -n) \big)-w_- \big(\prob(S_\tau \leq -n-1 ) \big)\Big) ,
\end{aligned}
\end{equation}
where $u_\pm(x)$ denote the utility function in the gain region and the disutility function in the loss region, respectively. Note that $u(x) = u_+(x) {\bf 1}_{x \ge 0} - u_-(-x) {\bf 1}_{x < 0}$.

Now we define the set of admissible stopping strategies, $\cal T$.
First, we require any stopping time $\tau\in \cal T$ to be such that the stopped process $(S_{\tau \wedge t}, t \in \bN)$ is uniformly integrable. This, among other things, is to rule out the notorious ``doubling strategies'' in the infinite time horizon.
Second, we include randomized strategies. As pointed out in \cite{HeHuOblojZhou2016:Randomization}, a randomized strategy can possibly strictly increase the CPT value $V$ because quasi-convexity does not hold. In other words, the gambler makes exit decision based on the information both from the gain/loss process and from some external source such as coin flipping.
Mathematically, let $({\cal F}_t, t \in \bN)$ denote the natural filtration generated by the asymmetric random walk $(S_t, t \in \bN)$.
The probability space is large enough to support a uniform random variable $\xi$ which is independent from $(S_t, t \in \bN)$.
Then there exist a sequence of mutually independent Bernoulli random variables $\xi_t^{S_j, j \le t}$, which are independent of the underlying gain/loss process $(S_t, t \in \bN)$. Here $\xi_t^{S_j, j \le t}=0$ if a coin toss lands on head and $\xi_t^{S_j, j \le t}=1$ if otherwise, at node $(t,S_t)$ with respect to the gain/loss history $S_j, j \le t$. In other words, to use which $\xi_t^{\cdot}$ at node $(t,S_t)$ depends on the history $(S_j,j \le t)$.
Let ${\cal G}_t := {\cal F}_t \vee \sigma(\xi_s^{S_j, j \le s}, s \le t)$ be an enlarged filtration which contains the original natural filtration and a $\sigma$-algebra generated by the Bernoulli random variables.
An admissible (randomized) stopping time $\tau\in \cal T$ is therefore ${\cal G}_t$-adapted.
We remark here that the probability distributions of Bernoulli random variables $\xi_t^{S_j, j \le t}$ are not exogenously given, but {\it derived} as a {\it part} of the optimal solution.
In other words, the gambler endogenously chooses the probability of heads or tail of coin, i.e., $\prob(\xi_t^{S_j, j \le t} = 1) = 1 - \prob(\xi_t^{S_j, j \le t} = 0)$.

The above formulated problem is time-inconsistent due to the presence of probability weighting function. As known in the time-inconsistent literature (e.g.  \citealp{BjorkMurgoci2014:TheoryMarkovianTimeInconsistent}), there are three type of agents when facing time-inconsistency. First, the precomitted agent solves the problem at time 0 and follows through the optimal solution at later times. Second, the na\"ive one solves the problem at each time--state pair and ``pastes" the resulting solutions together. Third, the sophisticated agent considers equilibrium strategies such that the decisions made at each time--state node are consistent with the equilibrium.
This paper will focus on the precommitted type in tackling  our problem.

\section{Skorokhod embedding}\label{se:Skorokhodembedding}

Due to time-inconsistency,
conventional techniques such as martingale approach and dynamic programming fail to solve the optimal stopping problem \eqref{optprob}. Here we employ the Skorokhod embedding, which was also used to solve the fair gamble problem in \cite{HeEtal2019:StoppingStrategies}.
Because the objective function \eqref{objtfunc} is law-invariant in $S_\tau$, we first change the decision variable from stopping time $\tau$ to the probability distribution function of $S_\tau$. If we can characterize the set of admissible probability distribution functions, then we will be able to turn \eqref{optprob} into an infinite-dimensional program on the distribution of $S_\tau$.
Finally, we recover the optimal $\tau$ from the optimal distribution of $S_\tau$.
Therefore, the key step is to characterize the set of the probability distributions of $S_\tau$
where $\tau\in \cal T$. Such a result is called a Skorokhod embedding theorem.

We first establish the uniform integrability of $(S_{t \wedge \tau}, t \in \bN)$.
The following lemma shows that $(S_{\tau \wedge t}, t \in \bN)$ is uniformly integrable if and only if $S_{\tau}$ is $\mathcal{L}^1$-integrable.

\begin{lemma}\label{le:uniformintegrable}
$(S_{t \wedge \tau}, t \in \bN)$ is uniformly integrable if and only if $S_{\tau}$ is $\mathcal{L}^1$-integrable, that is, $\mathbb{E} |S_{\tau}| < +\infty$. In this case, we have $\mathbb{E}[\tau] = \mathbb{E}[S_\tau] / (2p-1)$.
\end{lemma}

Lemma \ref{le:uniformintegrable} shows that the uniform integrability of $(S_{t \wedge \tau}, t \in \bN)$ is identical to the $\mathcal{L}^1$ integrability of $S_\tau$, which in turns depends on the latter's probability distribution. This property is necessary to characterize feasible probability measures that can embed the asymmetric random walk.
Moreover, the expectation of the stopping time can be obtained. Also note that this lemma collapses for the symmetric case when $p=1/2$.

In the unfair gambles described above, the probability of winning one dollar is strictly smaller than that of losing one dollar in each round. Because $p < 1/2$ the gain/loss process $(S_t, t \in \bN)$ is a super-martingale.
This non-martingale underlying process makes it difficult to apply directly the two Skorokhod embeddings employed in the fair gamble problem in \cite{HeHuOblojZhou2017:Twoexplicitembedding} that relies on  changing the decision variable from stopping times to probability distribution functions.
This is because applying the optional sampling theorem to a super-martingale  only gives inequalities, not equalities.
To overcome this difficulty, we consider a new process by transforming the asymmetric random walk $(S_t, t \in \bN)$ to a martingale.
Define a function $g_\rho(x) := \rho^x - 1$, where $\rho := (1-p)/p > 1$.
Let $\widetilde S_t := g_\rho(S_t)$. Because the function $g_\rho$ is strictly increasing, values of $\widetilde S_t$ have one-to-one correspondence to values of $S_t$.
However, the new process $(\widetilde S_t, t \in \bN)$ is a martingale with respect to the filtration $({ \cal F}_t, t \in \bN)$ because
\begin{align*}
\mathbb{E}[\widetilde S_{t+1} | \mathcal{F}_t] = \mathbb{E}[\rho^{S_{t+1}} -1 | \mathcal{F}_t] = \rho^{S_t + 1} p + \rho^{S_t -1 } (1-p) -1 = \rho^{S_t} -1 = \widetilde S_t \;.
\end{align*}

Denote by $\mathcal{M}^\rho$ the set of probability measures on the integer set $\mathbb{Z}$ that can be embedded into the asymmetric random walk $(S_t, t \in \bN)$ with the odds ratio $\rho$, namely, for any $\mu \in \mathcal{M}^\rho$, there exists some ${\cal G}_t$-adapted randomized stopping time $\tau$ such that $S_\tau$ has the same probability distribution as $\mu$.
The following proposition gives the necessary conditions for a probability measure to be embedded into the asymmetric random walk $(S_t, t \in \bN)$ (henceforth termed ``feasible").

\begin{proposition}\label{prop:necessarycond}
We have $\mathcal{M}^\rho \subseteq \mathcal{\widetilde M}^\rho$ where 
\begin{align*}
\mathcal{\widetilde M}^\rho := \left\{\mu : \sum_{y \in {\mathbb{Z}}} \mu(\{y\}) = 1, \; \sum_{y \in {\mathbb{Z}}} |y| \mu(\{y\}) < \infty, \; \sum_{y \in {\mathbb{Z}}} g_\rho(y) \mu(\{y\}) \le 0 \right\}.
\end{align*}
\end{proposition}

According to Proposition \ref{prop:necessarycond}, any feasible  probability measure $\mu$ satisfies the conditions specified in the bracket on the right hand side above.
Note that the uniform integrability of $(\widetilde S_{\tau \wedge t}, t \in \mathbb{N})$ is not necessary for feasibility of $(S_t, t \in \bN)$. Indeed, if $(\widetilde S_{\tau \wedge t}, t \in \mathbb{N})$ is uniformly integrable, then
\begin{align*}
\sum_{y \in \mathbb{Z}} g_\rho(y) \mu(\{y\}) = \mathbb{E}[\widetilde S_{\tau}] = 0 .
\end{align*}
Conversely,  if $\mathbb{E}[\widetilde S_{\tau}] = 0$, then $\lim_{t \to \infty} \mathbb{E}[\widetilde S_{\tau \wedge t} + 1] = 1 = \mathbb{E}[\widetilde S_{\tau} + 1]$.
Since $\widetilde S_{\tau \wedge t} \to \widetilde S_{\tau} $, a.s., and $\widetilde S_{\tau \wedge t} + 1 \geq 0$ for all $t$, by Scheffe's lemma, we have $\lim_{t \to \infty} \mathbb{E}[ |\widetilde S_{\tau \wedge t} - \widetilde S_{\tau}| ] = 0$ and hence, $(\widetilde S_{\tau \wedge t}, t \in \bN)$ is uniformly integrable.

The following result, however,  shows that the uniform integrability of $(\widetilde S_{\tau \wedge t}, t \in \bN)$ is a {\it sufficient} condition of a feasible probability measure $\mu$.

\begin{proposition}\label{prop:sufficientcond}
We have $\mathcal{M}^\rho \supseteq \mathcal{\widehat M}^\rho$ where 
\begin{align*}
 \mathcal{\widehat M}^\rho := \left\{\mu : \sum_{y \in {\mathbb{Z}}} \mu(\{y\}) = 1, \; \sum_{y \in {\mathbb{Z}}} |y| \mu(\{y\}) < \infty, \; \sum_{y \in {\mathbb{Z}}} g_\rho(y) \mu(\{y\}) = 0 \right\}.
\end{align*}
Moreover, for $\mu \in \mathcal{\widehat M}^\rho$, there exists a randomized Az{\'e}ma--Yor stopping time $\tau_{AY}$ such that $S_{\tau_{AY}} \sim \mu$.
\end{proposition}

Proposition \ref{prop:sufficientcond} is proved based on an explicit construction of randomized Az\'ema--Yor stopping times.
We illustrate how randomized Az\'ema-Yor stopping times is constructed below.
Given $\mu \in \mathcal{\widehat M}^\rho$, define left-continuous function $\psi_{\mu}$ to be
\begin{align}\label{de:psi}
\psi_{\mu}(x) := \frac{1}{\sum_{g_\rho(k) \geq x, k \in \bZ} \mu(\{k\})} \sum_{g_\rho(k) \geq x, k \in \bZ } g_\rho(k) \mu(\{k\}),
\end{align}
and $b_{\mu}$ the right-continuous inverse of $\psi_{\mu}$:
\begin{align}\label{de:b}
b_{\mu}(x) := \min \left\{ g_\rho(k) :  \psi_{ \mu}(g_\rho(k)) \leq x \text{ and } \psi_{\mu}(g_\rho(k)+) > x, k \in \mathbb{Z} \right\}.
\end{align}
For any $n \in \bN$, let $x^n_1 > ... > x^n_{m_n} > x^n_{m_n+1}$ be the corresponding atoms of $b_{\mu}$ on the interval $[g_\rho(n), g_\rho(n+1)]$:
\begin{align}\label{de:x}
x^n_1 = b_{\mu}(g_\rho(n+1)), ... , x^n_{m_n} = b_{\mu}(g_\rho(n)+),x^n_{m_n+1} = b_{\mu}(g_\rho(n)),
\end{align}
where $m_n+1$ is the total number of the atoms.

Suppose that the gain/loss process $(S_t, t\in\bN)$ reaches a new running maximum level $n\in\bN$ at some time prior to stopping. Because $S_t$ and $\widetilde S_t$ has one-to-one correspondence, it is equivalent  to stopping the process $(\widetilde S_t, t \in \bN)$ which arrives at the maximum $g_\rho(n)$. This leads to the following stopping strategy in  three steps:

{\em Step 1}: Set the upper bound to be $g_\rho(n+1)$ and lower bound to be $x^n_{1}$. If the upper bound is reached first after $u$ steps, then the maximum is updated to be $\widetilde S_{t+u} = g_\rho(n+1)$. If the lower bound $x^n_{1}$ is reached first, go to next step.

{\em Step 2}: Toss a coin with specified probabilities of tail/head (to be determined). If it shows a tail, stop; if it shows a head, continue with the lower bound reset to be $ x^n_{2}$ and without changing the upper bound, which is $g_\rho(n+1)$. This procedure continues until either the process attains the new running maximum $g_\rho(n+1)$, or by tossing a coin the gambler stops at $x^n_k \in \mathbb{Z}$, $k = 1,2,...m_n$, or if he reaches the last lower bound labeled $x^n_{m_n+1} \in \mathbb{Z}$, then go to next step.

{\em Step 3}: Stop for sure once $x^n_{m_n+1}$ is reached.

In words, our randomized Az\'ema--Yor type stopping strategy works as follows.
Whenever the gain/loss process reaches a new maximum value, if the gambler has not stopped yet, the stopping region is updated; the gambler stops for sure if the gap between the current cumulative gain/loss state and the running maximum exceeds a certain level; he  tosses a coin to decide whether to stop or continue if the gap is an intermediate value, where the probability of heads or tails is endogenously determined by the probability distribution of the cumulative gains or loss at that time; and he continues for sure if the gap is small.
Therefore, such stopping times are path-dependent.
In a later section we will present concrete examples of those stopping strategies.

\section{Infinite-dimensional programming}\label{se:indp}
Based on the Skorokhod embedding established in the previous section, we can change the original optimal stopping problem to an infinite-dimensional program.
The objective function \eqref{objtfunc} is law-invariant so we can change the decision variable from the stopping time $\tau$ to the probability distribution function of $S_\tau$. In other words, the optimal stopping problem \eqref{optprob} is equivalent to be written as
\begin{equation*}
\begin{array}{cl}
\underset{\mu \in \mathcal{M}^\rho} {\text{sup }} V(\mu),
\end{array}
\end{equation*}
where $\mu$ is the probability measure corresponding to $S_\tau$ and thus,
\begin{align*}
V(\mu) = & \sum_{n=1}^\infty u_+(n) \Big(w_+ \big( \sum_{j = n}^\infty \mu(\{j\}) \big)-w_+ \big( \sum_{j = n+1}^\infty \mu(\{j\}) \big) \Big) \\
 & \quad - \sum_{n=1}^\infty u_-(n) \Big(w_- \big(\sum_{j = n}^\infty \mu(\{-j\}) \big)- w_- \big( \sum_{j = n+1}^\infty \mu(\{-j\}) \big) \Big).
\end{align*}

We now further specify the objective and the constraints.
Given a probability measure $\mu$, let $z_{+,n} := \sum_{j = n}^\infty \mu(\{j\})$, and $z_{-,n} := \sum_{j = n}^\infty \mu(\{-j\})$, $n \in \bN_+$.
Then naturally 
\begin{align}
1\geq{z_{+,1}}\geq{z_{+,2}}\geq{...}\geq{z_{+,n}}\geq{...}\geq0, \label{eq:zpluscond} \\
1\geq{z_{-,1}}\geq{z_{-,2}}\geq{...}\geq{z_{-,n}}\geq{...}\geq0, \label{eq:zminuscond} \\
z_{+,1} + z_{-,1} \leq 1. \label{eq:z1cond}
\end{align}
Denote by ${\bf z_{+,-}} = (z_{+,1},...,z_{+,n},...;z_{-,1},...,z_{-,n},...)$. Let
\begin{align*}
v({\bf z_{+,-}}) : = \sum_{n=1}^\infty u_+(n) \Big(w_+(z_{+,n})-w_+(z_{+,n+1})\Big) - \sum_{n=1}^\infty u_-(n) \Big(w_-(z_{-,n})-w_-(z_{-,n+1})\Big).
\end{align*}
It is straightforward to see the one-to-one correspondence between the probability measure $\mu$ and the vector ${\bf z_{+,-}}$ with $V(\mu) = v({\bf z_{+,-}})$.

For a probability measure $\mu \in \mathcal{\widetilde M}^\rho$, we further have the $\mathcal{L}^1$-integrability condition
\begin{align}\label{eq:znormcond}
\sum_{n=1}^\infty z_{+,n} + \sum_{n=1}^\infty z_{-,n} = \sum_{j \in \bZ} |j| \mu(\{j\}) < \infty,
\end{align}
and the inequality
\begin{align}
0 \geq \sum_{j \in {\mathbb{Z}}} g_\rho(j) \mu(\{j\}) & = \sum_{j = 1}^\infty \left( \rho^j -1 \right) \mu(\{j\}) + \sum_{j=1}^\infty \left( \rho^{-j} -1 \right) \mu(\{-j\}) \notag \\
& = (1-2p) \left( \frac{1}{1-p} \sum_{n=1}^\infty \rho^{n} z_{+,n} - \frac{1}{p} \sum_{n=1}^\infty \rho^{-n} z_{-,n} \right). \label{eq:zcond}
\end{align}

Consider a relaxed optimal stopping problem with feasible set $\mathcal{\widetilde M}^\rho$:
\begin{equation}\label{prob:relaxed}
\begin{array}{cl}
\underset{\mu \in \mathcal{\widetilde M}^\rho} {\text{sup }} V(\mu). \tag{$\rm \tilde P$}
\end{array}
\end{equation}
This relaxed problem is equivalent to an infinite-dimensional program with decision variable ${\bf z_{+,-}}$:
\begin{equation}\label{prob:UnfairGames:infinitedim}
\begin{array}{cl}
\underset{ {\bf z_{+,-}} }{\text{sup }} & v({\bf z_{+,-}}), \\
\text{subject to  } & \eqref{eq:zpluscond} - \eqref{eq:zcond}. \tag{InP}
\end{array}
\end{equation}
Denote by ${\bf z}^*_{+,-}$ the optimal solution to \eqref{prob:UnfairGames:infinitedim} if it exists.
The following theorem shows that for optimal ${\bf z}^*_{+,-}$, the equality of the last constraint \eqref{eq:zcond} holds. It then implies that the corresponding probability measure $\mu^*$ must belong to $\mathcal{\widehat M}^\rho$ as well as $\mathcal{M}^\rho$. Hence, the optimal probability distribution can be embedded into the corresponding asymmetric random walk.
\begin{theorem}\label{thm:UnfairGames:UI}
Suppose the optimal solution ${\bf z}^*_{+,-}$ to the infinite-dimensional programming problem \eqref{prob:UnfairGames:infinitedim} exists, and $\mu^*$ is the corresponding probability measure solving the relaxed optimization problem \eqref{prob:relaxed}. Then $\mu^* \in \mathcal{\widehat M}^\rho$. Hence, there exists a randomized stopping time $\tau^* \in {\cal T}$ such that $S_{\tau^*}$ has the same probability distribution as $\mu^*$ and $(S_{\tau^* \wedge t}, t \in \mathbb{N})$ is uniformly integrable. Moreover,
\begin{align*}
\mathbb{E}[\tau^*] = \frac{1}{2p-1}\left(\sum_{n=1}^\infty z_{+,n}^* - \sum_{n=1}^\infty z_{-,n}^*\right).
\end{align*}
\end{theorem}

According to Theorem \ref{thm:UnfairGames:UI}, if ${\bf z}^*_{+,-}$ is the optimal solution to problem \eqref{prob:UnfairGames:infinitedim}, then from the probability distribution we can recover the stopping strategy through randomized Az\'ema--Yor stopping time which solves the original optimal stopping problem \eqref{optprob}.
Henceforth we investigate problem \eqref{prob:UnfairGames:infinitedim}.

\cite{HeEtal2019:StoppingStrategies} show that in the fair gambling model, the value of the optimal stopping strategy is infinite for ceratin parameter values.
By contrast, in the current unfair gambling problem where $p \in (0,1/2)$, the gambler is expected to lose $|p-(1-p)| = 1-2p$ dollar in each round of bet; so  the longer he stays in the casino the more he is expected to lose. Hence, conditions under which the fair gambling problem has infinite objective value may no longer lead to an infinite objective value in the unfair counterpart.
Indeed, we will show that the infinite-dimensional program \eqref{prob:UnfairGames:infinitedim} has finite objective values for a wide range of parameter specifications.

We adopt the convention that $v({\bf z_{+,-}}):=-\infty$ if the second sum of infinite series in $v({\bf z_{+,-}})$ is infinite.

\begin{proposition}\label{prop:wellposedness}
Suppose that $|u_+(n)-u_+(n-1)|$ is bounded for $n \in \bN$.
If the probability weighting function $w_+(\cdot)$ satisfies
\begin{align}\label{eq:UnfairGames:wellposedness}
\limsup_{n \to \infty} \left|  w_+\left( \frac{1}{\rho^n-1} \right) \right|^{\frac{1}{n}} < 1,
\end{align}
then problem (\ref{prob:UnfairGames:infinitedim}) has a finite objective value.
\end{proposition}

Note that the boundedness of $|u_+(n)-u_+(n-1)|$ is easily satisfied by commonly used utility functions, including piece-wise power S-shaped utility.
One can also verify that the condition \eqref{eq:UnfairGames:wellposedness} holds for the following probability weighting functions.
First, consider the power concave probability weighting function, i.e., $w_+(p) = p^{\delta_+}$, $\delta_+ \in (0,1)$. Then
\begin{align*}
\limsup_{n \to \infty} \left| w_+\left( \frac{1}{\rho^n-1} \right) \right|^{\frac{1}{n}} = \lim_{n \to \infty} \left| \left( \frac{1}{\rho^n-1} \right)^{\delta_+} \right|^{\frac{1}{n}}  = \left( \frac{1}{\rho} \right)^{\delta_+} < 1.
\end{align*}
Second, consider the probability weighting function proposed by \cite{TverskyKahneman1992:CPT}, i.e.,
\begin{align*}
w_+(p) = \frac{p^{\delta_+}}{(p^{\delta_+}+(1-p)^{\delta_+})^{\frac{1}{\delta_+}}}, \; \delta_+ \in (0,1).
\end{align*}
Since $(p^{\delta_+}+(1-p)^{\delta_+})^{\frac{1}{\delta_+}} \ge 1$ for $p \in [0,1]$, $w_+(p) \le p^{\delta_+}$.
\begin{align*}
\limsup_{n \to \infty} \left| w_+\left( \frac{1}{\rho^n-1} \right) \right|^{\frac{1}{n}} \leq \limsup_{n \to \infty} \left|\left( \frac{1}{\rho^n-1} \right)^{\delta_+}\right|^{\frac{1}{n}} = \left( \frac{1}{\rho} \right)^{\delta_+} < 1.
\end{align*}
Third, consider the probability weighting function proposed by \citet{GoldsteinEinborn1987:ExpressionTheory}, i.e.,
\begin{align*}
w_+(p) = \frac{b_+ p^{\delta_+}}{b_+ p^{\delta_+} +(1-p)^{\delta_+}}, \; b_+ > 0, \delta_+ \in (0,1).
\end{align*}
Then
\begin{align*}
\limsup_{n \to \infty} \left| w_+\left( \frac{1}{\rho^n-1} \right) \right|^{\frac{1}{n}} = \lim_{n \to \infty} \left( \frac{b_+}{b_+ + (\rho^n-2)^{\delta_+}} \right) ^{\frac{1}{n}} = \left( \frac{1}{\rho} \right)^{\delta_+} < 1.
\end{align*}
Therefore, problem \eqref{prob:UnfairGames:infinitedim} has finite objective values under the above probability weighting functions.
On the other hand, for the probability distortion function proposed by \cite{PrelecD:98df}, i.e., $w_+(p) = e^{-{(-\log p)}^{\frac{1}{2}}}$, condition \eqref{eq:UnfairGames:wellposedness} does not hold and \eqref{prob:UnfairGames:infinitedim} may then be ill-posed.

\section{Analytic solutions}\label{se:powerutilityanddistortion}

Following \cite{HeEtal2019:StoppingStrategies}, in this section, we solve the infinite-dimensional program \eqref{prob:UnfairGames:infinitedim} for piece-wise power utility function and power probability weighting functions, that is,
\begin{align}\label{eq:powerfunction}
u_+(x) = x^{\alpha_+}, \; u_-(x) = \lambda x^{\alpha_-}, \; 0< \alpha_\pm < 1, \; \lambda \ge 1; \quad w_\pm(p) = p^{\delta_\pm}, \; 0 < \delta_\pm \le 1.
\end{align}
The problem \eqref{prob:UnfairGames:infinitedim} is  decomposed into three sub-problems: gain-part problem, loss-part problem, and merged problem.
Recall that the decision variable in \eqref{prob:UnfairGames:infinitedim} is ${\bf z_{+,-}} = (z_{+,1},...,z_{+,n},...;z_{-,1},...,z_{-,n},...)$.

\subsection{Gain-part problem}\label{subse:Gainpartsubproblem}
Fix $z_{+,1}$ and denote $z_n := z_{+,n+1}/z_{+,1}$, $n \in \bZ_+$. Let ${\bf z} := (z_1,z_2,...)$. The gain-part problem is defined as follows:
\begin{equation}\label{prob:UnfairGames:GainPart}
\begin{array}{cl}
  \underset{{\bf z}}{\text{sup }} &v_+({\bf z};s):=\sum_{n=1}^\infty \Big((n+1)^{\alpha_+}-n^{\alpha_+}\Big)z_n^{\delta_+}\\
  \text{subject to  }  &1\geq{z_1}\geq{z_2}\geq{...}\geq{z_n}\geq{...}\geq0,\\
  &\sum_{n=1}^\infty \rho^{n} {z_n}=s. \tag{GP}
\end{array}
\end{equation}

\begin{proposition}\label{prop:UnfairGames:GainPart}
For $j \in \bZ_+$, let
\begin{align*}
A_j := \sum_{n=j}^\infty \left(\frac{1}{\rho^n} \Big((n+1)^{\alpha_+}-n^{\alpha_+}\Big)\right)^\frac{1}{1-\delta_+}, \;
W_j := \sum_{n=j}^\infty \left(\frac{1}{\rho^{n \delta_+}} \Big((n+1)^{\alpha_+}-n^{\alpha_+}\Big) \right)^\frac{1}{1-\delta_+}.
\end{align*}
\begin{enumerate}
\item[(i)]
If $0 \le s < A_1 \left(\frac{1}{\rho}(2^{\alpha_+}-1) \right)^\frac{1}{\delta_+-1}$, then the optimal solution to \eqref{prob:UnfairGames:GainPart} is
\begin{align*}
z_n^* = s A_1^{-1} \left( \frac{1}{\rho^n} \Big((n+1)^{\alpha_+}-n^{\alpha_+}\Big)\right)^{\frac{1}{1-\delta_+}} , \quad n=1,2,...
\end{align*}
The optimal value is
$v_{+}^*(s) = {s}^{\delta_+}{A_1}^{-\delta_+} W_1$.
\item[(ii)]
If $s \ge A_1 \left(\frac{1}{\rho}(2^{\alpha_+}-1) \right)^\frac{1}{\delta_+-1}$ and 
let $j \in \bZ_+$ be such that
\begin{align*}
j-1 + A_{j} \left( \frac{1}{\rho^j} ( (j+1)^{\alpha_+}-{j}^{\alpha_+}) \right)^\frac{1}{\delta_+-1} \le s < j + A_{j+1} \left( \frac{1}{\rho^{j+1}} ({(j+2)}^{\alpha_+}-{(j+1)}^{\alpha_+}) \right)^\frac{1}{\delta_+-1},
\end{align*}
 then the optimal solution to \eqref{prob:UnfairGames:GainPart} is
\begin{align*}
z_n^*=
\begin{cases}
1 &, n = 1,...,j \\
(s-j) A_{j+1}^{-1} \left( \frac{1}{\rho^n} \Big((n+1)^{\alpha_+}-n^{\alpha_+}\Big)\right)^{\frac{1}{1-\delta_+}}  &, n=j+1,...
\end{cases}
\end{align*}
The optimal value is
$v_{+}^*(s) = (j+1)^{\alpha_+}-1 + (s-j)^{\delta_+} {A_{j+1}}^{-\delta_+} W_{j+1}$.
\end{enumerate}
\end{proposition}

The following proposition shows that the optimal value has the same order as $(\log s)^{\alpha_+}$.
\begin{proposition}\label{prop:UnfairGames:GainPart:valuefunction}
For sufficiently large values of $s$, there exist positive constants $c_{+}$ and $C_{+}$ such that
\begin{align*}
c_{+} \left(\log {s}\right)^{\alpha_+} < v_{+}^*(s) < C_{+} \left(\log {s}\right)^{\alpha_+}.
\end{align*}
\end{proposition}

\subsection{Loss-part problem}\label{subsubse:Losspartsubproblem}
Fix $z_{-,1}$ and again denote $z_n := z_{-,n+1}/z_{-,1}$. Let ${\bf z} = (z_1,z_2,...)$. For simplicity, let $\phi := \rho^{-1} < 1$. The loss-part problem is defined as follows:
\begin{equation}\label{prob:UnfairGames:LossPart}
\begin{array}{cl}
  \underset{{\bf z}}{\text{inf }} & v_-({\bf z};s) := \sum_{n=1}^\infty \Big((n+1)^{\alpha_-}-n^{\alpha_-}\Big)z_n^{\delta_-}\\
  \text{subject to  }  &1\geq{z_1}\geq{z_2}\geq{...}\geq{z_n}\geq{...}\geq0, \; \sum_{n=1}^\infty z_n < +\infty, \\
  &\sum_{n=1}^\infty \phi^{n} z_n = s. \tag{LP}
\end{array}
\end{equation}
Let $\mathcal{Z}_-(s) :=  \{{\bf z} = (z_1,z_2,...): 1\geq{z_1}\geq{z_2}\geq{...}\geq0, \sum_{n=1}^\infty z_n < +\infty \text{ and } \sum_{n=1}^\infty \phi^n{z_n}=s \}$, which is the set of feasible solution to \eqref{prob:UnfairGames:LossPart}.

Note that $s < \phi/(1-\phi)$.
First, let us suppose $0 \le s < \phi$.
Consider the feasible solution to \eqref{prob:UnfairGames:LossPart}, ${\bf z}^{0,k} = (z^{0,k}_1, z^{0,k}_2, ...)$ for $k \in \bZ_+$, with the following form:
\begin{align*}
z^{0,k}_1 = z^{0,k}_2 = ... = z^{0,k}_k = \frac{s(1-\phi)}{\phi(1-\phi^k)}, \quad z^{0,k}_{k+1} = z^{0,k}_{k+2} = ... = 0.
\end{align*}
Let $\mathcal{Z}_-^0(s) := \{{\bf z}^{0,k}, k \in \bZ_+ \}$.

\begin{proposition}\label{prop:UnfairGames:infinites01}
Suppose $0 \le s < \phi$. Then any feasible solution ${\bf z} \in \mathcal{Z}_-(s)$ is a convex combination of ${\bf z}^{0,k} \in \mathcal{Z}_-^0(s)$, $k \in \bZ_+$. Therefore, the optimal solution to Problem (\ref{prob:UnfairGames:LossPart}) belongs to $\mathcal{Z}_-^0(s)$, i.e., ${\bf z}^* \in \mathcal{Z}_-^0(s)$.
\end{proposition}

Now in general, suppose $\phi (1-\phi^{m-1})/(1-\phi) \le s < \phi (1-\phi^{m})/(1-\phi)$ for some $m \in \bZ_+$. Consider the feasible solution to \eqref{prob:UnfairGames:LossPart}, ${\bf z}^{j,k} = (z^{j,k}_1, z^{j,k}_2,...)$ for $0 \le j \le m-1$ and $k \ge m$, with the following form:
\begin{align*}
z^{j,k}_1 = z^{j,k}_2 = ... = z^{j,k}_j= 1, \quad z^{j,k}_{j+1} = ... =  z^{j,k}_{k} = \frac {s(1-\phi)-{\phi (1-\phi^{j})}}{{\phi^{j+1} (1-\phi^{k-j})}}, \quad z^{j,k}_{k+1} = ... = 0.
\end{align*}
For each $j \in [0,m-1] \cap \bZ$, let $\mathcal{Z}_-^j(s) := \{{\bf z}^{j,k}, k \ge m \}$.

\begin{proposition}\label{prop:UnfairGames:infinitesmminus1m}
Suppose $\phi (1-\phi^{m-1})/(1-\phi) \le s < \phi (1-\phi^{m})/(1-\phi)$ for some $m \in \bZ_+$.
Then any feasible solution ${\bf z} \in \mathcal{Z}_-(s)$ is a convex combination of ${\bf z}^{j,k} \in \mathcal{Z}_-^j(s)$, $0 \le j \le m-1$, $k \ge m$.
Therefore, the optimal solution to Problem (\ref{prob:UnfairGames:LossPart}) belongs to $\cup_{j=0}^{m-1} \mathcal{Z}_-^j(s)$, i.e., ${\bf z}^* \in \cup_{j=0}^{m-1} \mathcal{Z}_-^j(s)$.
\end{proposition}

Note that given $s \geq 0$, $\phi (1-\phi^{m-1}) / (1-\phi) \leq s < \phi (1-\phi^{m})/(1-\phi)$,  implying 
\begin{align*}
m = m(s) := \left\lfloor \frac{\log(\phi-s(1-\phi)) }{\log \phi } \right \rfloor \geq 1,
\end{align*}
which is integer-valued. By Proposition \ref{prop:UnfairGames:infinitesmminus1m}, solving Problem \eqref{prob:UnfairGames:LossPart} is equivalent  to solving the following optimization problem:
\begin{equation*}
\begin{array}{cl}
  \underset{j,k}{\text{inf }} &R_m(j,k) := v_-({\bf z}^{j,k};s) \\
  \text{subject to  } & j = 0,1,...m-1, \; k = m, m+1,...,
\end{array}
\end{equation*}
where
\begin{align*}
 v_-({\bf z}^{j,k};s) = \Big( (j + 1)^{\alpha_-} - 1 \Big) + \Big( (k+1)^{\alpha_-}-(j+1)^{\alpha_-} \Big) \left( \frac {s{(1-\phi)}-{\phi (1-\phi^{j})}}{{\phi^{j+1} (1-\phi^{k-j})}} \right)^{\delta_-}.
\end{align*}

We have the following results.
\begin{proposition}\label{prop:UnfairGames:losspart:optimall1}
\begin{enumerate}
\item[(i)]  Suppose $\alpha_- \geq \delta_-$. Then for fixed $m \ge 1$ and $j \in [0,m-1] \cap \bZ$, $R_m(j,k)$ is  minimized at $k = m$.
\item[(ii)] Suppose $\alpha_- < \delta_-$.
\begin{enumerate}
\item[(a)] If $\phi \leq e^{-\delta_-(1-\alpha_-)}$, then for fixed $m \ge 1$ and $j \in [0,m-1] \cap \bZ$, $R_m(j,k)$ is minimized at $k = m$.
\item[(b)] If $\phi > e^{-\delta_-(1-\alpha_-)}$, then for fixed $m \ge \lceil \delta_-(1-\alpha_-)/(-\log\phi) \rceil$ and $j \in [\delta_-(1-\alpha_-)/(-\log\phi) -1,m-1] \cap \bZ$, $R_m(j,k)$ is minimized at $k = m$.
\item[(c)] If $\phi > e^{-\delta_-(1-\alpha_-)}$, then for fixed $m \ge \lceil \delta_-(1-\alpha_-)/(-\log\phi) \rceil$ and $j \in [0, \delta_-(1-\alpha_-)/(-\log\phi) -1) \cap \bZ$ or $m < \lceil \delta_-(1-\alpha_-)/(-\log\phi) \rceil$ and $j \in [0,m-1] \cap \bZ$, define
    \begin{align}\label{eq:gmj}
    g_{m,j}(x) := \frac{ \phi^{-x}-1 }{ -\log \phi} - \frac{\delta_-}{\alpha_-}  \left( (j+1+x)-(j+1)^{\alpha_-}(j+1+x)^{1-\alpha_-} \right).
    \end{align}
    If there exists $x_0 > 0$ such that $g_{m,j}(x_0) = 0$, $g_{m,j}(x_0-) < 0$ and $g_{m,j}(x_0+) > 0$, then
\begin{enumerate}
\item if $j + \lfloor x_0 \rfloor \ge m+1$,  $R_m(j,k)$ is minimized at $k \in \{m, j+\lfloor x_0 \rfloor, j + \lfloor x_0 \rfloor + 1\}$;
\item if $j + \lfloor x_0 \rfloor = m$,  $R_m(j,k)$ is minimized at $k \in \{m, m + 1\}$;
\item if $j + \lfloor x_0 \rfloor \le m-1$,  $R_m(j,k)$ is minimized at $k = m$.
\end{enumerate}
Otherwise if there exists no such $x_0$, then $R_m(j,k)$ is minimized at $k = m$.
\end{enumerate}
\end{enumerate}
\end{proposition}


\subsection{Merged problem}\label{subse:Mergingproblem}
Combining the gain-part  and the loss-part problems leads to  the following ``merged problem":
\begin{equation}\label{prob:UnfairGames:OptimalProbabilityDistribution}
\begin{array}{cl}
\underset{z_{+,1},z_{-,1},s_+,s_-}{\text{sup }} &v_M(z_{+,1},z_{-,1},s_+,s_-)=(v_{+}^*(s_+)+1)(z_{+,1})^{\delta_+} -\lambda (v_{-}^*(s_-)+1)(z_{-,1})^{\delta_-}\\
\text{subject to  }
& 0 \le z_{+,1} \le 1, \; 0 \le z_{-,1} \le 1, \; z_{+,1} + z_{-,1} \le 1, \\
& 0 \le s_+ < +\infty, \; 0 \le s_- < \frac{\phi}{1-\phi}, \\
& (s_++1)z_{+,1} = \phi (s_-+1)z_{-,1}. \tag{MP}
\end{array}
\end{equation}

To solve \eqref{prob:UnfairGames:OptimalProbabilityDistribution}, for given $s_+$ and $s_-$, let
\begin{align*}
f(y) := (v_{+}^*(s_+)+1) \phi^{\delta_+} \left(\frac{s_-+1}{s_++1} \right)^{\delta_+} y^{\delta_+} -\lambda (v_{-}^*(s_-)+1) y^{\delta_-}\;.
\end{align*}

\begin{proposition}\label{prop:mergedproblem}
\begin{enumerate}
\item[(i)] Suppose $\delta_+ \ge \delta_-$. Let
\begin{align*}
M := \sup_{0 \le s_+ < +\infty, 0 \le s_- < \frac{\phi}{1-\phi}} \frac{(v_{+}^*(s_+)+1) \phi^{\delta_+} \left({s_-+1} \right)^{\delta_+}}{(v_-^*(s_-)+1)\left({{s_++1}} \right)^{\delta_-} (s_++1+\phi(s_-+1))^{\delta_+-\delta-}} \;.
\end{align*}
\begin{enumerate}
\item[(a)] If $\lambda \ge M$, the optimal solution to \eqref{prob:UnfairGames:OptimalProbabilityDistribution} is $z_{+,1}^* = z_{-,1}^* = 0$, meaning that the gambler does not enter the game at all.
\item[(b)] If $\lambda < M$, the gambler will enter the game. The optimal solution to \eqref{prob:UnfairGames:OptimalProbabilityDistribution} is
\begin{align}\label{eq:sol1}
\begin{cases}
(s_+^*, s_-^*) = \underset{0 \le s_+ < +\infty, 0 \le s_- < \frac{\phi}{1-\phi}}{\text{argsup}}f \left(\frac{s_++1}{s_++1+\phi(s_-+1)} \right), \\
z_{+,1}^* = \frac{\phi (s_-^*+1)}{s_+^*+1+\phi(s_-^*+1)}, \; z_{-,1}^* = \frac{s_+^*+1}{s_+^*+1+\phi(s_-^*+1)}\;.
\end{cases}
\end{align}
\end{enumerate}
\item[(ii)] Suppose $\delta_+ < \delta_-$. The gambler will enter the game.
The optimal solution to \eqref{prob:UnfairGames:OptimalProbabilityDistribution} is
\begin{align}\label{eq:sol2}
\begin{cases}
(s_+^*, s_-^*) = \underset{0 \le s_+ < +\infty, 0 \le s_- < \frac{\phi}{1-\phi}}{\text{argsup}} f(k(s_+,s_-)), \\
z_{+,1}^* = \phi\frac{s_-^*+1}{s_+^*+1} k(s_+^*,s_-^*), \; z_{-,1}^* = k(s_+^*,s_-^*) \;.
\end{cases}
\end{align}
where
\begin{align*}
k(s_+,s_-) = \min \left\{\left( \frac{{\delta_+} (v_{+}^*(s_+)+1) \phi^{\delta_+}\left({s_-+1} \right)^{\delta_+}}{\lambda {\delta_-} (v_{-}^*(s_-)+1) \left({s_++1} \right)^{\delta_+} } \right)^{\frac{1}{\delta_--\delta_+}}, \frac{s_++1}{s_++1+\phi(s_-+1)} \right\} \;.
\end{align*}
\end{enumerate}
\end{proposition}

\section{Comparison to fair gambles}\label{se:UnfairGamesNumericalExample}
In this subsection we present numerical examples of optimal stopping strategies in the unfair gambling problem. Following the previous analysis, we first derive the optimal probability distribution of cumulative gains or loss at exit time and then recover the stopping strategy in the form of randomized Az{\'e}ma--Yor stopping time. The results are finally compared to the fair gambling counterparts. 

\begin{figure}[!htbp]
  \centering
  \begin{minipage}[t]{0.3\textwidth}
    \includegraphics[width=\textwidth]{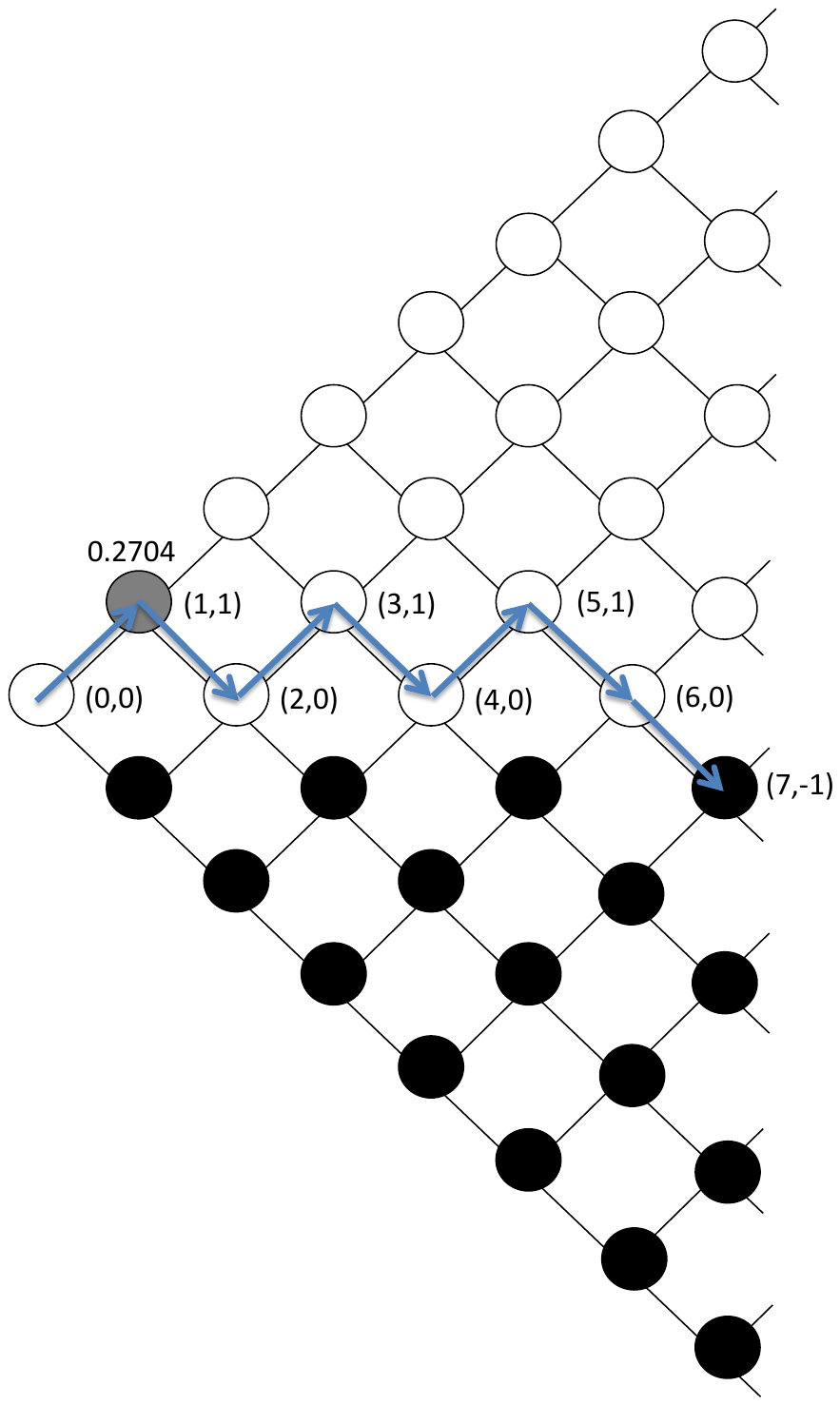}
  \end{minipage}
  \begin{minipage}[t]{0.3\textwidth}
    \includegraphics[width=\textwidth]{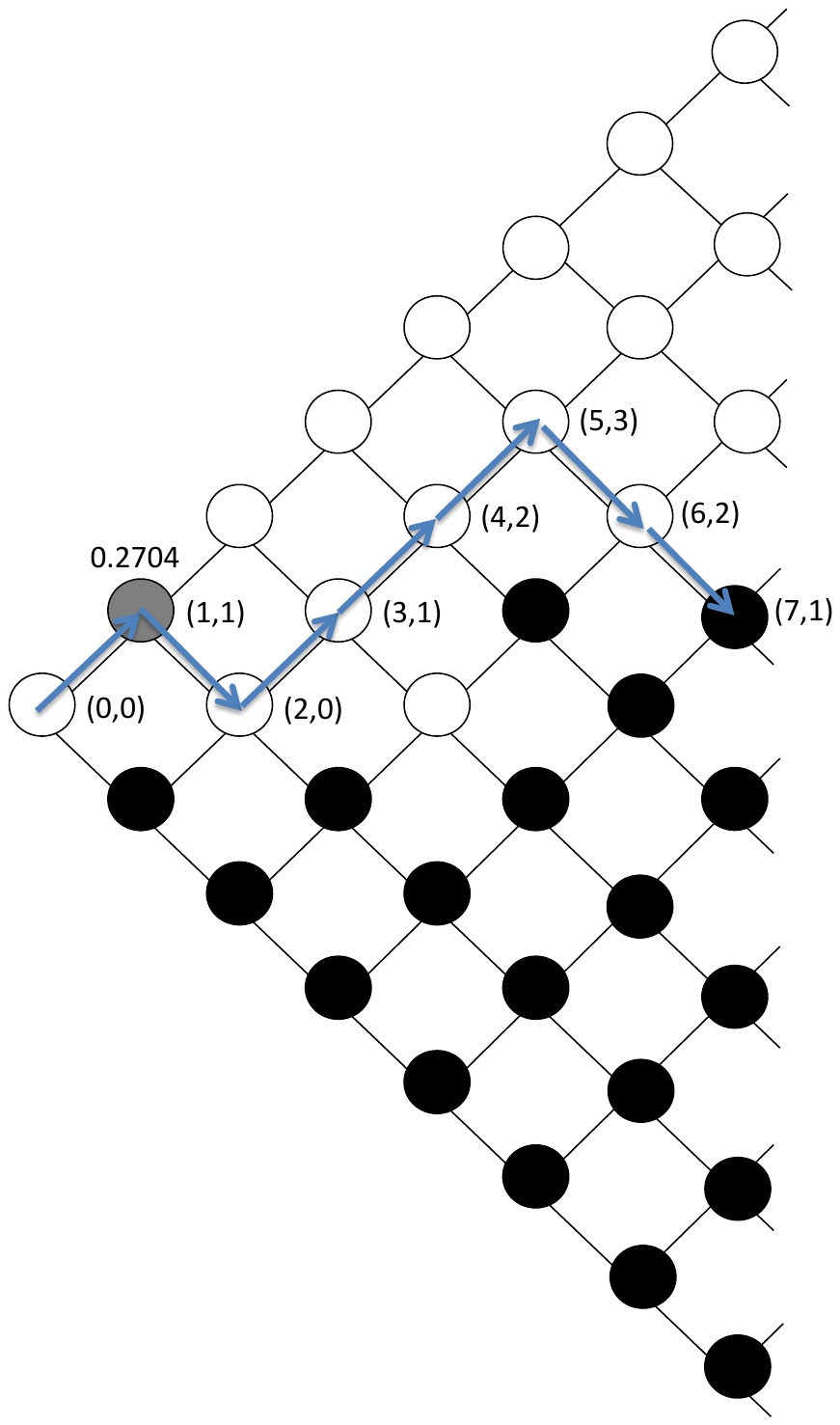}
  \end{minipage}
  \begin{minipage}[t]{0.3\textwidth}
  \includegraphics[width=\textwidth]{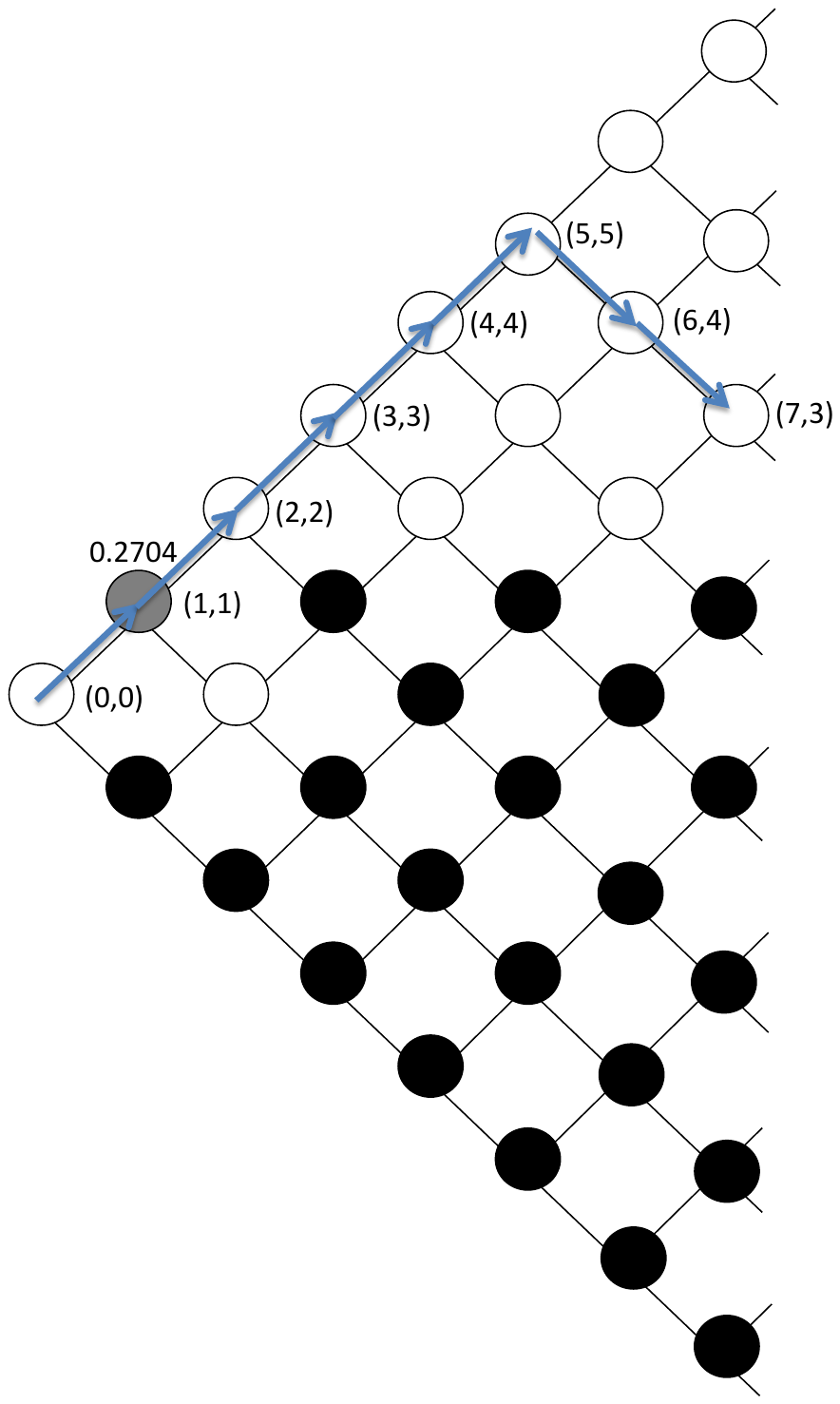}\\
  \end{minipage}
  \caption{Precommitted strategies in symmetric gambles in three paths}\label{fig:StoppingStrategyforfairGames}
\end{figure}

Suppose the utility function is piece-wise power and probability distortion functions are power, which are given in \eqref{eq:powerfunction}.
Let $\alpha_+ = 0.6$, $\delta_+ = 0.7$, $\alpha_- = 0.8$, $\delta_- = 0.7$, and $\lambda = 1.05$, which is used in  \cite{HeEtal2019:StoppingStrategies}.
First, we recall the optimal solution in symmetric gambles in \cite{HeEtal2019:StoppingStrategies} as a benchmark case. The optimal CPT value is 0.0632, meaning that the precommitted gambler will enter the fair game at time 0.
The optimal distribution of stopped gains or loss is
\begin{align*}
\begin{cases}
p_j^* = 0.4465\left(\left(j^{0.6}-(j-1)^{0.6}\right)^{\frac{1}{0.3}} - \left((j+1)^{0.6}-j^{0.6}\right)^{\frac{1}{0.3}}\right),\;\; j \ge 2, \\
p_1^* = 0.3297, \; p_0 = 0, \; p_{-1}^* = 0.6216, \\
p_{-2}^* =p_{-3}^* =...=0.
\end{cases}
\end{align*}
Then the stopping strategy is recovered from the optimal distribution and illustrated in Figure \ref{fig:StoppingStrategyforfairGames}. Each node is colored in either white, black, or grey: if the node is white, it means continue; if the node is black, it means stop; if the node is grey, it means randomization, with the number above the node standing for the probability of tossing a tail (stop).
Because it is a randomized, path-dependent strategy, we use three different paths of gain/loss process to illustrate the stopping strategy taken by the gambler. Note that due to the path-dependence property of the strategy, the nodes colors are updated in real time according to its path.

Figure \ref{fig:StoppingStrategyforfairGames} left panel shows one path. The gambler starts at node $(0,0)$ with the lower bound equal to $-1$. First, he arrives at node $(1,1)$, which is grey. Hence, he tosses a coin with the probability of tail (stop) equal to $0.2704$. Suppose it shows a head so he continues the game. Note that this path does not achieve a new maximum before the lower bound $-1$ is hit. He finally stops at a black node $(7,-1)$ with a loss equal to $-1$.
Now, consider the second path in Figure \ref{fig:StoppingStrategyforfairGames} middle panel. It is same as the first path until time $3$. At time $4$, the gambler arrives at node $(4,2)$. The new maximal gains of $2$ is achieved and afterwards, the lower stopping bound is no longer $-1$ but updated to $1$. He finally stops at node $(7,1)$ with a gain equal to $1$.
Finally, consider the third path in Figure \ref{fig:StoppingStrategyforfairGames} right panel. In this path the gambler continues winning until node $(5,5)$ then loses in the next two steps. When the running maximum achieves $5$, the lower bound becomes $2$ so the gambler does not stop at node $(7,3)$.

Now consider $p = 0.49$, which is slightly smaller than $0.5$. Then $\phi = 49/51$. In this case the gambler accepts the bet. The optimal solution to \eqref{prob:UnfairGames:OptimalProbabilityDistribution} is $s_+^* = 0.333$, $s_-^* = 0$, $z_{+,1}^* = 0.4189$, $z_{-,1}^* = 0.5811$. The optimal CPT value is $0.006026$, which is much smaller than that of the fair game. The optimal probability distribution of cumulative gains or loss at exit time is
\begin{align*}
\begin{cases}
p_j^* = 0.6537 \left( \left(\frac{49}{51}\right)^{\frac{j-1}{0.3}} (j^{0.6}-(j-1)^{0.6})^{\frac{1}{0.3}}-\left(\frac{49}{51}\right)^{\frac{j}{0.3}}((j+1)^{0.6}-j^{0.6})^{\frac{1}{0.3}}\right), \;\;j \ge 2, \\
p_1^* = 0.3560, \; p_0 = 0, \; p_{-1}^* = 0.5811, \\
p_{-2}^* =p_{-3}^* =...=0.
\end{cases}
\end{align*}
By Theorem \ref{thm:UnfairGames:UI}, there exists a stopping time $\tau$ such that $\mathbb{P}(S_\tau = n) = p_n^*$, $n \in \mathbb{Z}$, and $S_{\tau \wedge n}$ is uniformly integrable.
Moreover, $\expect[\tau^*] = 5.34$, which indicates that in such an asymmetric game, on average, the gambler stops in five to six rounds.

\begin{figure}[!htbp]
  \centering
  \begin{minipage}[t]{0.3\textwidth}
  \includegraphics[width=\textwidth]{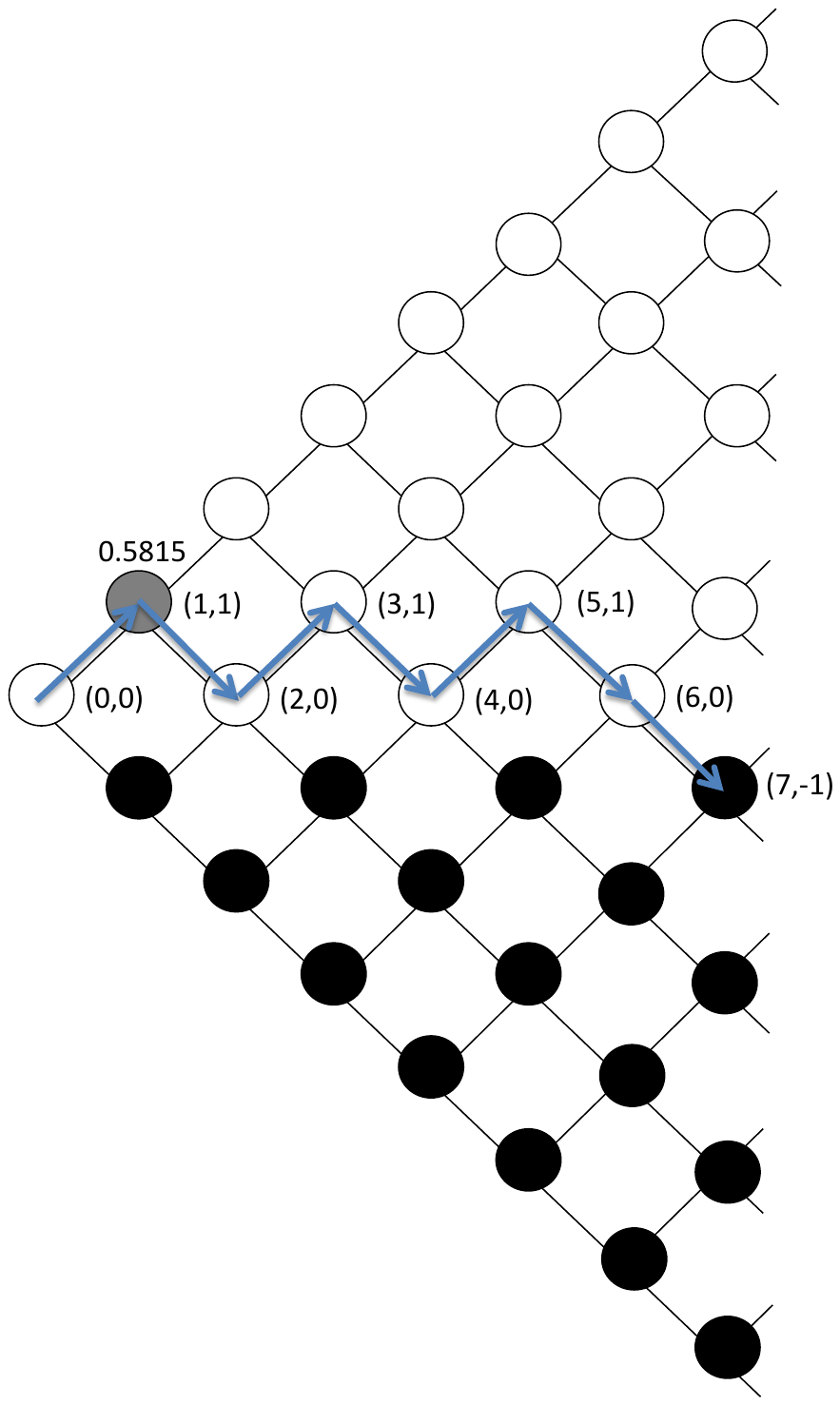}
  \end{minipage}
  \begin{minipage}[t]{0.3\textwidth}
  \includegraphics[width=\textwidth]{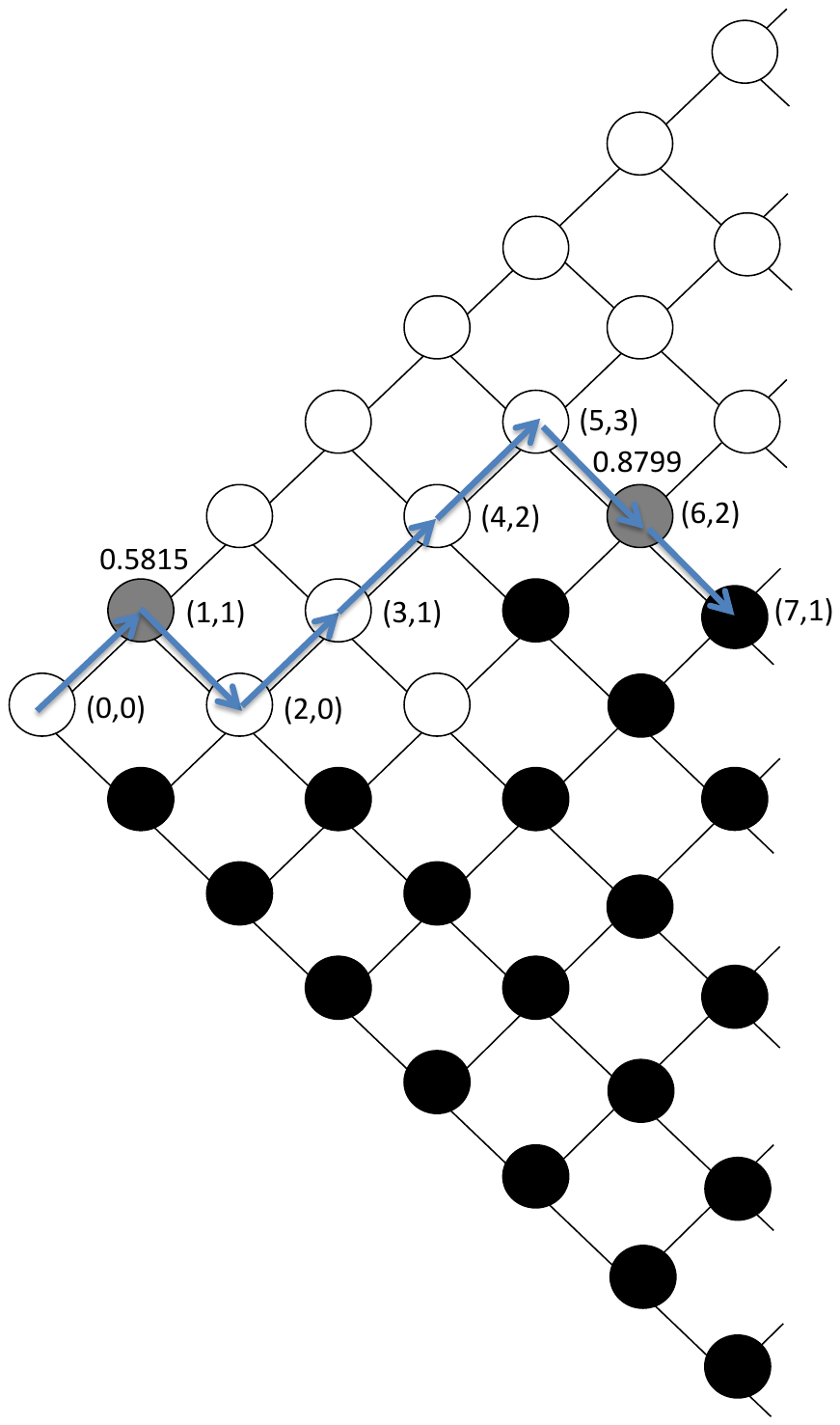}\\
  \end{minipage}
  \begin{minipage}[t]{0.3\textwidth}
  \includegraphics[width=\textwidth]{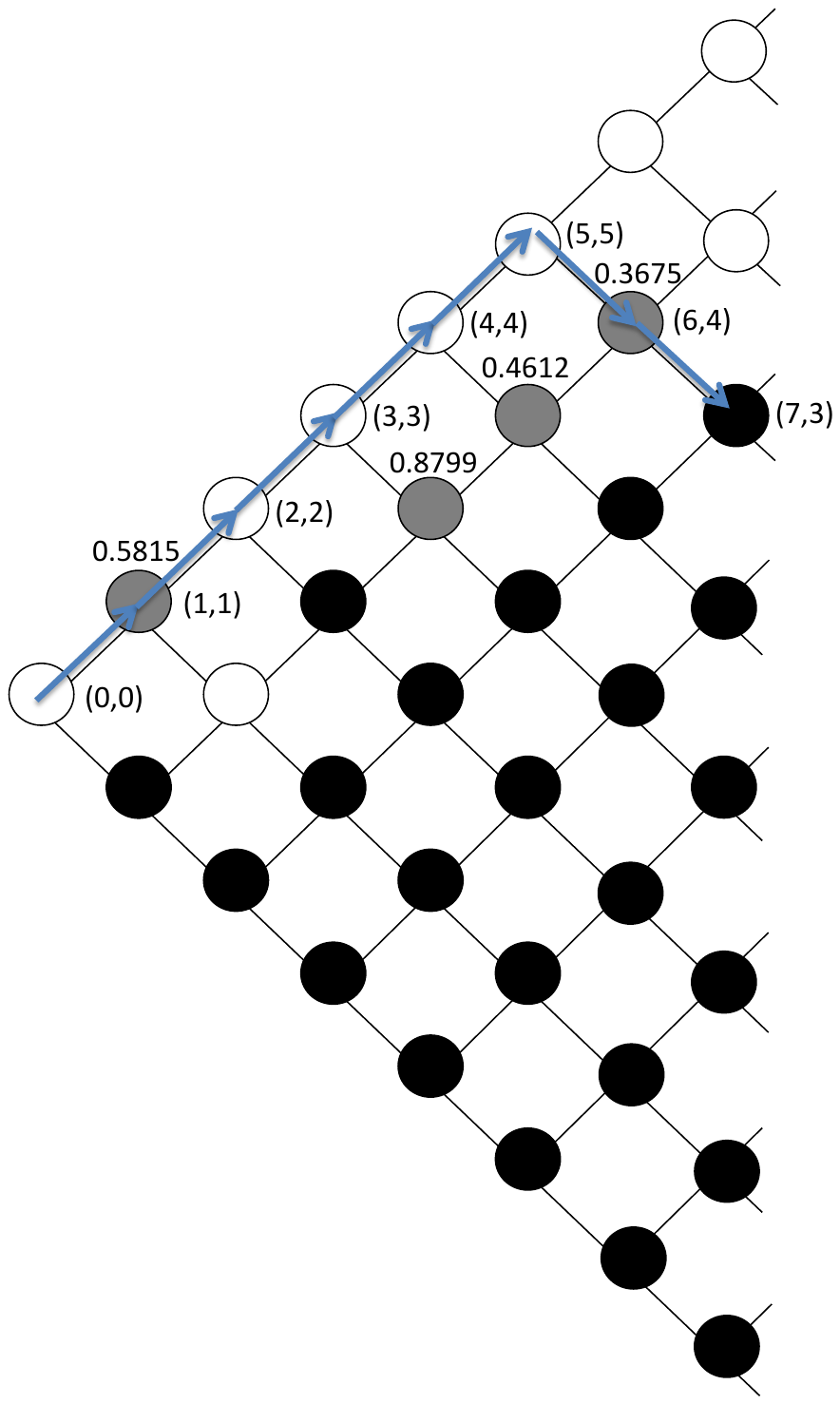}\\
  \end{minipage}
  \caption{Precommitted strategies in asymmetric gambles in three paths}\label{fig:StoppingStrategyforUnfairGames}
\end{figure}

The stopping strategy is illustrated with three paths corresponding to those in Figure \ref{fig:StoppingStrategyforfairGames}. Again, due to the path-dependence property of the strategy, the nodes colors are updated in real time according to the path. Figure \ref{fig:StoppingStrategyforUnfairGames} left panel shows one path.
Note that this path is same as the one in the fair game example shown in Figure \ref{fig:StoppingStrategyforfairGames} left panel. Both two strategies show that node $(1,1)$ is grey. However, the two strategies differ in the probability of tail at node $(1,1)$. The probability of stopping at node $(1,1)$ by tossing a coin is 0.5815, much larger than that in the  symmetric game.
For the second path shown in Figure \ref{fig:StoppingStrategyforUnfairGames} middle panel, in addition to the difference at node $(1,1)$ between symmetric and asymmetric games, node $(6,2)$ in the asymmetric game is grey with probability of stopping equal to 0.8799, while the one in the symmetric game is white.
Now, consider the third path illustrated in Figure \ref{fig:StoppingStrategyforUnfairGames} right panel.
The gambler starts at node $(0,0)$ with the lower exit bound equal to $-1$. Then he arrives at grey node $(1,1)$ and tosses a coin with the probability of tail (stop) equal to 
0.5815. Suppose it shows a head so he continues the game. He then continues winning until reaching node $(5,5)$ with the lower bound updated to $3$. Then he goes to node $(6,4)$, which is grey, and he tosses a coin with the probability of tail equal to 0.3675. Suppose it shows a head so he continues the game. Then he reaches the black node $(7,3)$ and finally stops.
In contrast to Figure \ref{fig:StoppingStrategyforfairGames} right panel, the gambler exits the asymmetric game earlier and is less loss-tolerant due to the unfavorable odds of the bet.

Keep the other parameter values unchanged and let $p = 1/3$ so that $\phi = p/(1-p) = 0.5$. Then the optimal solution to \eqref{prob:UnfairGames:OptimalProbabilityDistribution} is $z_{+,1}^* = 0$, $z_{-,1}^* = 0$, which means that the gambler simply dismisses such an unfair game at the outset. 
Thus the game is not attractive even to a CPT gambler when $p$ is sufficiently small.
This pattern also appears under other parameter values. Take 
$\alpha_\pm = 0.88, \delta_+ = 0.61, \delta_+ = 0.69, \lambda = 2.25$ for example, which are \cite{TverskyKahneman1992:CPT} estimates, the gambler will play in the fair game and the optimal CPT value is infinite. However, in an unfair game with the same preference parameters but $p<0.3563$ , the gambler will not enter the casino.

\section{Concluding Remarks}\label{se:UnfairGamesConcludingRemarks}
In this paper, we solve the behavioral gambling problem for unfair games. 
We restore martingality essential for the Skorokhod embedding technique to work by applying 
a geometric transformation to the underlying asymmetric random walk. In contrast to fair gambles, the problem has finite objective values with well-known probability weighting functions. If the probability of winning at each play is sufficiently small, the gambler simply does not enter the game. When he does, his stopping strategy is more conservative than that in the fair game for the same parameter setting. In particular, for the same gambling history the probability to stop at the same node is higher in the unfair game than in the fair game. As a result, the gambler tends to leave earlier in an unfair gamble and thus is less loss-tolerant.
Finally, we note that insights from the study of casino gambling may help understand agents' behaviors in other more practical and important settings such as stock markets and venture capital investment.

\bibliography{LongTitles,BibFile}

\bigskip

\appendix

\noindent{\Large \bf Appendix}

\setcounter{equation}{0}

\renewcommand\theequation{A.\arabic{equation}}

\section{Proof}

\pfof{Lemma \ref{le:uniformintegrable}}
This proof follows the idea in \cite{GranditsFalker2000}[Proposition 2.2].
We first show the necessary part.
If $(S_{t \wedge \tau}, t \in \bN)$ is uniformly integrable, then $\forall \epsilon >0$, $\exists K > 0$ large enough such that $\mathbb{E}[ |S_{t \wedge \tau} | {\bf 1}_{|S_{t \wedge \tau}| > K} ] < \epsilon$ for all $t \in \mathbb{N}$. By Fatou's lemma, we have
\begin{align*}
\mathbb{E} |S_{\tau}| \leq \liminf_{t \to \infty} \mathbb{E} |S_{t \wedge \tau}| = \liminf_{t \to \infty} \left\{ \mathbb{E}[ |S_{t \wedge \tau}| {\bf 1}_{|S_{t \wedge \tau}| \leq K}] + \mathbb{E}[ |S_{t \wedge \tau}| {\bf 1}_{|S_{t \wedge \tau}| > K}] \right\} \\
\leq K + \sup_{t \in \mathbb{N}} \mathbb{E}[ |S_{t \wedge \tau}| {\bf 1}_{|S_{t \wedge \tau}| > K} ] < K + \epsilon < +\infty.
\end{align*}
Hence, $S_{\tau}$ is $\mathcal{L}^1$-integrable.

We next show the sufficient part. Suppose that $\mathbb{E} |S_{\tau}| < +\infty$.
Let $Y_t := S_t + (1-2p)t/2$ and $\bar Y:= \sup_{t \in \mathbb{N}} Y_t$. Then for any $t \in \mathbb{N}$,
\begin{align}\label{eq:propUnfairGamesUI:martingale1}
S_t = Y_t - \frac{(1-2p)t}{2} \leq \bar Y - \frac{(1-2p)t}{2} .
\end{align}
We are going to show that $\mathbb{E} |\bar Y| < +\infty$.
Define a function
\begin{align*}
f(x) := px^{1 + (1-2p)/2} + (1-p)x^{- 1 + (1-2p)/2} -1, \; x > 0.
\end{align*}
Then the first order-derivative of $f$ is
\begin{align*}
f'(x) & =  p \left(1 + \frac{1-2p}{2} \right) x^{(1-2p)/2} + (1-p) \left(- 1 + \frac{1-2p}{2} \right) x^{- 2 + (1-2p)/2} \\
& = x^{(1-2p)/2} \left( \frac{p(3-2p)}{2} + \frac{(1-p)(-1-2p)}{2} x^{- 2} \right).
\end{align*}
It is straightforward to verify that $f$ is first decreasing on $(0, \bar x)$ then increasing on $(\bar x, \infty)$, where $\bar x > 1$ is such that $f^{'}(\bar x) = 0$. Since $f(1) = 0$, there exists $\kappa > \bar x$ such that $f(\kappa) = 0$.
Hence, $(\kappa^{Y_t}, t \in \bN)$ is a martingale because
\begin{align*}
\mathbb{E}[ \kappa^{Y_{t+1}} | \mathcal{F}_n ] & = p\kappa^{S_t + 1 + \frac{1-2p}{2}(t+1)} + (1-p)\kappa^{S_t - 1 + \frac{1-2p}{2}(t+1)} \\
& = \kappa^{Y_t} \left( p\kappa^{1 + (1-2p)/2} + (1-p)\kappa^{- 1 + (1-2p)/2} \right) = \kappa^{Y_t}.
\end{align*}
Let $\tau_{A,B} := \inf \{t \in \mathbb{N} : Y_t \geq A \text{ or } Y_t \leq -B \}$, where $A, B > 0$. By optional sampling theorem, we have
\begin{align*}
\mathbb{E}[ \kappa^{X_{\tau_{A,B}}} ] = p(A)\kappa^{A} + (1-p(A)) \kappa^{-B} = 1,
\end{align*}
where $p(A)$ is the probability of $Y_t$ reaching $A$ before $-B$. Then
\begin{align*}
p(A) = \frac{1-\kappa^{-B}}{\kappa^{A}-\kappa^{-B}} \;.
\end{align*}
Letting $B \to \infty$, we have
\begin{align*}
\mathbb{P}(\bar Y \geq A) = \mathbb{P}(\sup_t Y_t \geq A) = \mathbb{P}(Y_t \text{ ever reaches } A) = \frac{1}{\kappa^{A}}.
\end{align*}
It is then straightforward to verify that $\mathbb{E} |\bar Y| < \infty$. Hence, by (\ref{eq:propUnfairGamesUI:martingale1}) we have
\begin{align}\label{eq:propUnfairGamesUI:tau}
\mathbb{E}[\tau] \leq \frac{2}{1-2p} \left(\mathbb{E} [Y] - \mathbb{E} [S_{\tau}]\right) < \infty.
\end{align}

Now, we show that $(S_{\tau \wedge t}, t \in \mathbb{N})$ is dominated by $\mathcal{L}^1$-integrable random variable. Let $\hat S_t := S_t^2-(1-(1-2p)^2) t$. Then $(\hat S_t, t \in \mathbb{N})$ is a martingale because
\begin{align*}
\expect[\hat S_{t+1} | \cF_t] = &\mathbb{E}[ S_{t+1}^2 -(1-(1-2p)^2) (t+1) | \mathcal{F}_t ] \\
=& p(S_t + 1 + (1-2p))^2 + (1-p)(S_t - 1 + (1-2p))^2 -(1-(1-2p)^2)(t+1)\\
=& S_t^2 - (1-(1-2p)^2) t = \hat S_t.
\end{align*}
Then $\expect[S_{t \wedge \tau}^2] = (1-(1-2p)^2) \expect[t \wedge \tau]$.
By Doob's maximal inequality,
\begin{align*}
\mathbb{E}[\sup_t S_{\tau \wedge t}^2] \leq 4 \liminf_{t \to \infty} \mathbb{E}[S_{\tau \wedge t}^2] = 4((1-(1-2p)^2)  \lim_{t \to \infty} \mathbb{E}[\tau \wedge t] = 4((1-(1-2p)^2) \mathbb{E}[\tau] < \infty,
\end{align*}
where the last inequality follows from (\ref{eq:propUnfairGamesUI:tau}). Hence, we have
\begin{align*}
\mathbb{E}[\sup_t | S_{\tau \wedge t} |] \leq \left( \mathbb{E}[\sup_t S_{\tau \wedge t}^2] \right)^{\frac{1}{2}}  < \infty.
\end{align*}
Therefore, $(S_{\tau \wedge t}, t \in \mathbb{N})$ is dominated by $\sup_t | S_{\tau \wedge t}|$ which is $\mathcal{L}^1$-integrable, and $(S_{\tau \wedge t}, t \in \mathbb{N})$ is uniformly integrable.

Finally, note that $(S_t - (2p-1)t, t \in \bN)$ is a martingale:
\begin{align*}
\expect[S_{t+1} - (2p-1)(t+1) | \cF_t] = (S_t + 1)p + (S_t - 1)(1-p) - (2p-1)(t+1) \\
= S_t - (2p-1)t.
\end{align*}
Then $\expect[S_{t \wedge \tau}] = (2p-1) \expect[t \wedge \tau]$.
By dominated convergence theorem,
\begin{align*}
\expect[S_\tau] = \lim_{t \to \infty} \expect[S_{t \wedge \tau}] = (2p-1) \lim_{t \to \infty} \expect[t \wedge \tau] = (2p-1) \expect[\tau],
\end{align*}
where the last equality follows from monotone convergence theorem. Hence, $\mathbb{E}[\tau] = \frac{1}{2p-1}\mathbb{E}[S_\tau]$.\halmos

\pfof{Proposition \ref{prop:necessarycond}}
Given $\mu \in \mathcal{M}^\rho$, we have $\sum_{y \in {\mathbb{Z}}} \mu(\{y\}) = 1$ since $\mu$ is a probability measure taking values on integers, and there exists a randomized stopping time $\tau$ such that $S_\tau \sim \mu$ and $(S_{t \wedge \tau}, t \in \bN)$ is uniformly integrable. By Lemma \ref{le:uniformintegrable}, $\expect|S_\tau| < \infty$. Therefore, $\sum_{x \in {\mathbb{Z}}} |y| \mu(\{y\}) < \infty$.

Recall the definition of $(\widetilde S_t, t \in \bN)$.
Then $\mathbb{E}[\widetilde S_{\tau \wedge t}] = \mathbb{E}[\widetilde S_{0}] = 0$ for any $t \in \mathbb{N}$.
Hence,
\begin{align*}
\sum_{y \in \mathbb{Z}} g_\rho(y) \mu(\{y\}) = \mathbb{E}[\widetilde S_{\tau}] \leq \liminf_{t \to \infty} \mathbb{E}[\widetilde S_{\tau \wedge t} ] = 0 \;,
\end{align*}
where the inequality follows from Fatou's lemma and $\widetilde S_{\tau \wedge t} > -1$ for all $t \in \mathbb{N}$.
As a result, $\mu \in \mathcal{\widetilde M}^\rho$.
\halmos

\pfof{Proposition \ref{prop:sufficientcond}}
Given $\mu \in \mathcal{\widehat M}^\rho$, denote by $p_k := \mu(\{k\})$ for $k \in \bZ$. Recall the definition of $\psi_\mu$, $b_\mu$, and $x^n_1,...x^n_{m_n+1}$ as in \eqref{de:psi} -- \eqref{de:x}.

Denote by $\mathcal{H}_{g_\rho(n)}$ the hitting time of $(S_t, t \in \bN)$ onto the state $n$ which is also the hitting time of $(\widetilde S_t, t \in \bN)$ onto the state $g_\rho(n)$:
\begin{align*}
\mathcal{H}_{n} = \inf \{ t \in \mathbb{N} : S_t \geq n \} = \inf \{ t \in \mathbb{N} : \widetilde S_t \geq {g_\rho(n)} \}.
\end{align*}
Let $\mu_\rho(\{x\}) := \mu(\{g_\rho^{-1}(x)\})$, ${\bar \mu}_\rho(x) := {\bar \mu}(g_\rho^{-1}(x)) := \sum_{g_\rho(k) \geq x, k \in \bZ} p_k$.
We claim that for $j \in \bN$,
\begin{align}\label{eq:AYlikestoppingtime}
\begin{cases}
\mathbb{P}(\widetilde S_{\tau} = x \text{ and } \tau < \mathcal{H}_{j+1}) = \mu_\rho(\{x\}), &\text{ for } x < x^j_1,  \\
\mathbb{P}(\widetilde S_{\tau} = x \text{ and } \tau < \mathcal{H}_{j+1}) = {\bar \mu}_\rho(x^j_1) \frac{{g_\rho(j+1)}-\psi_{\mu}( x^j_1)}{{g_\rho(j+1)}-x^j_1},  &\text{ for } x = x^j_1, \\
\mathbb{P}(\widetilde S_{\tau} = x \text{ and } \tau < \mathcal{H}_{j+1}) = 0,  &\text{ for } x > x^j_1.
\end{cases}
\end{align}

We first verify that \eqref{eq:AYlikestoppingtime} holds for $j = 0$.
At time 0, according to the three-step stopping rule, first set the upper bound to be $g_\rho(1)$ and lower bound to be $x^0_{1}$. If the lower bound $x^0_{1}$ is reached, toss a coin $\xi^0_{1}$ with probability of tail set to be
\begin{align*}
\prob(\xi^0_1 = 1) = r^0_{1} := {\bar \mu}_\rho(x^0_{1}) \frac{{g_\rho(1)}-\psi_{\mu}( x^0_{1})}{{g_\rho(1)}-g_\rho(0)} \in [0,1].
\end{align*}
Then
\begin{align*}
\mathbb{P}(\widetilde S_{\tau} = x^0_{1} \text{ and } \tau < \mathcal{H}_{1}) & = \mathbb{P}(x^0_{1} \text{ is reached before } g_\rho(1) \text{ and } \xi^0_1 = 1) \\
& = \frac{g_\rho(1) - g_\rho(0)}{g_\rho(1) - x^0_{1}} r^0_{1} =  {\bar \mu}_\rho(x^0_{1}) \frac{{g_\rho(1)}-\psi_{\mu}( x^0_{1})}{{g_\rho(1)}-x^0_{1}} \;.
\end{align*}
If the coin shows a head, continue with upper bound $g_\rho(1)$ and updated lower bound, so on so forth; if the lower bound $x^0_{k}$ is reached, toss a coin $\xi^0_{k}$ with probability of tail set to be
\begin{align*}
\prob(\xi^0_k = 1) = r^0_{k} :&= \frac{\mu_\rho(\{x^0_{k}\}) (g_\rho(1) - x^0_{k})}{(1 - \bar \mu_\rho(x^0_{k-1})) g_\rho(1) + \sum_{g_\rho(y) \ge x^0_{k-1}, y \in \bZ} g_\rho(y) p_y} \\
& = \frac{\mu_\rho(\{x^0_{k}\}) (g_\rho(1) - x^0_{k})}{\mu_\rho(\{x^0_{k}\}) (g_\rho(1) - x^0_{k}) + (1 - \bar \mu_\rho(x^0_{k})) g_\rho(1) + \sum_{g_\rho(y) \ge x^0_{k}, y \in \bZ} g_\rho(y) p_y} \in [0,1].
\end{align*}
Then
\begin{align*}
\mathbb{P}(\widetilde S_{\tau} = x^0_{k} \text{ and } \tau < \mathcal{H}_{1}) & = \mathbb{P}(x^0_{k} \text{ is reached before } g_\rho(1) \text{ and } \xi^0_1 = \cdots = \xi^0_{k-1} = 0, \xi^0_k = 1)\\
& = \frac{g_\rho(1) - g_\rho(0)}{g_\rho(1) - x^0_{k}} (1 - r^0_{1}) \cdots (1-r^0_{k-1})  r^0_{k} = \mu_\rho( \{x^0_{k}\}) \;.
\end{align*}
Finally, note that if the support of $\mu$ has a lower bound $\underline y > -\infty$, then $m_0 > -\infty$ such that $x^0_{m_0+1} = g_\rho( \underline y)$. Since $\bar \mu_\rho (x^0_{m_0+1}) = 1$ and $\sum_{y \in z} g_\rho(y) \mu(\{y\}) = 0$, we obtain $r^0_{m_0+1} = 1$.

Suppose (\ref{eq:AYlikestoppingtime}) holds for $j \le n-1$. Then
\begin{align}
\mathbb{P}(\tau \geq \mathcal{H}_{n})
& = 1 - \mathbb{P}(\tau < \mathcal{H}_{n}) \nonumber \\
& = 1 - \sum_{y \in \bZ } \mathbb{P}(\widetilde S_{\tau} = g_\rho(y) \text{ and } \tau < \mathcal{H}_{n}) \nonumber \\
& = 1 - \sum_{x < x^{n-1}_1} \mu_\rho(\{x\}) -  {\bar \mu}_\rho(x^{n-1}_1) \frac{{g_\rho(n)}-\psi_{\mu}(x^{n-1}_1)}{{g_\rho(n)}-x^{n-1}_1} \nonumber \\
& = {\bar \mu}_\rho(x^{n-1}_1) \frac{\psi_{\mu}( x^{n-1}_1) - x^{n-1}_1}{{g_\rho(n)}-x^{n-1}_1} \label{eq:stoppingtimedistribution}
 \;.
\end{align}
We verify that (\ref{eq:AYlikestoppingtime}) also holds for $j = n$.

\noindent{\bf Case 1: $m_n = 0$}

In this case $x^n_1 = x^n_{m_n+1} = x^{n-1}_1$. Then conditional on $\tau \geq \mathcal{H}_{n}$,
\begin{align*}
\tau \wedge \mathcal{H}_{n+1} = \inf\Big\{t \geq \mathcal{H}_{n} : \widetilde S_t \in \{x^n_1,g_\rho(n+1)\}\Big\}.
\end{align*}
According the stopping rule described by previous three steps,
\begin{align}\label{eq:conditionaldistribution}
\mathbb{P}(\widetilde S_{\tau} = x^{n}_1 \text{ and } \tau < \mathcal{H}_{n+1} | \tau \geq \mathcal{H}_{n}) & = \prob(x^{n}_1 \text{ is reached before } g_\rho(n+1) | \tau \geq \mathcal{H}_{n}) \nonumber \\
& = \frac{g_\rho(n+1) - g_\rho(n)}{g_\rho(n+1)- x^{n}_1} \;.
\end{align}
Using \eqref{eq:stoppingtimedistribution} and \eqref{eq:conditionaldistribution}, we have
\begin{align*}
&\mathbb{P}(\widetilde S_{\tau} = x^{n}_1 \text{ and } \tau < \mathcal{H}_{n+1}) \\
=&\mathbb{P}(\widetilde S_{\tau} = x^{n}_1 \text{ and } \tau < \mathcal{H}_{n})+ \mathbb{P}(\widetilde S_{\tau} = x^{n}_1 \text{ and } \tau < \mathcal{H}_{n+1} | \tau \geq \mathcal{H}_{n}) \cdot \mathbb{P}(\tau \geq \mathcal{H}_{n})\\
=&\mathbb{P}(\widetilde S_{\tau} = x^{n-1}_1 \text{ and } \tau < \mathcal{H}_{n})+ \frac{g_\rho(n+1) - g_\rho(n)}{g_\rho(n+1)-x^{n}_1} \cdot \mathbb{P}(\tau \geq \mathcal{H}_{n})\\
=& \bar {\mu}_\rho(x^{n-1}_1) \frac{g_\rho(n)-\psi_{\mu}(x^{n-1}_1)}{g_\rho(n)-x^{n-1}_1}+\frac{g_\rho(n+1) - g_\rho(n)}{g_\rho(n+1)- x^{n}_1} \cdot \bar {\mu}_\rho(x^{n-1}_1) \frac{\psi_{\mu}(x^{n-1}_1)-x^{n-1}_1}{g_\rho(n)-x^{n-1}_1}\\
=&\bar {\mu}_\rho(x^{n}_1) \frac{g_\rho(n+1)-\psi_{\mu}(x^{n}_1)}{g_\rho(n+1) - x^{n}_1}.
\end{align*}
For the states other than $x^{n}_1$, note that they are not stopped during $\mathcal{H}_{n} \le \tau < \mathcal{H}_{n+1}$, then
\begin{align*}
\mathbb{P}(\widetilde S_{\tau} = x \text{ and } \tau < \mathcal{H}_{n+1}) & = \mathbb{P}(\widetilde S_{\tau} = x \text{ and } \tau < \mathcal{H}_{n}) \\
& =
\begin{cases}
\mu_\rho(\{x\}), & \text{ for } x < x^{n}_1 \\
0, & \text{ for } x > x^{n}_1.
\end{cases}
\end{align*}
We conclude that when $m_n = 0$, (\ref{eq:AYlikestoppingtime}) is satisfied for $j = n$.

\noindent{\bf Case 2: $m_n=1$}

In this case $x^n_1 > x^n_2 = x^n_{m_n+1} = x^{n-1}_1$. According to the stopping rule, once the process $(\widetilde S_t, t \in \bN)$ arrives a new maximum $g_\rho(n)$, set the upper bound to be $g_\rho(n+1)$ and the lower bound $x^n_1$. If $x^n_1$ is reached first, toss a coin $\xi^n_1$ with specified probability of tail/head. If it shows a tail, stop; if it shows a head, update the lower bound to be $x^n_2$ and continue. Once $x^n_2$ is reached, stop for sure. Then, conditional on $\tau \geq \mathcal{H}_{n}$,
\begin{align*}
\tau \wedge \mathcal{H}_{n+1} = \begin{cases}
\inf\Big\{t \geq \mathcal{H}_{n}: \widetilde S_t \in \{x^n_1,g_\rho(n+1)\}\Big\} \text{ if } \xi^n_1 = 1\\
\inf\Big\{t \geq \mathcal{H}_{n}: \widetilde S_t \in \{x^n_2,g_\rho(n+1)\}\Big\} \text{ if } \xi^n_1 = 0,
\end{cases}
\end{align*}
where $\mathbb{P}(\xi^n_1 = 1) = r^n_1$, the probability of tossing a tail at state $x^n_1$, to be determined.

By definition, note that
\begin{align}
\frac{\psi_{\mu}(x^n_2) - x^n_2}{\psi_{\mu}(x^n_1)-x^n_2} & = \frac{\sum_{g_\rho(k) \ge x^n_2} p_k g_\rho(k) / \bar {\mu}_\rho(x^n_2) - x^n_2}{ \sum_{g_\rho(k) \ge x^n_1} p_k g_\rho(k) / \bar {\mu}_\rho(x^n_1) - x^n_2}       \nonumber  \\
& = \frac{(\sum_{g_\rho(k) \ge x^n_2} p_k g_\rho(k) - \sum_{g_\rho(k) \ge x^n_2} p_k x^n_2) / \bar {\mu}_\rho(x^n_2)}{(\sum_{g_\rho(k) \ge x^n_1} p_k g_\rho(k) - \sum_{g_\rho(k) \ge x^n_1} p_k x^n_2) / \bar {\mu}_\rho(x^n_1)} \nonumber \\
& = \frac{\sum_{g_\rho(k) \ge x^n_2} p_k (g_\rho(k) - x^n_2) / \bar {\mu}_\rho(x^n_2)}{\sum_{g_\rho(k) \ge x^n_1} p_k (g_\rho(k) - x^n_2) / \bar {\mu}_\rho(x^n_1)} \nonumber \\
& =  \frac{\sum_{g_\rho(k) \ge x^n_1} p_k (g_\rho(k) - x^n_2) / \bar {\mu}_\rho(x^n_2)}{\sum_{g_\rho(k) \ge x^n_1} p_k (g_\rho(k) - x^n_2) / \bar {\mu}_\rho(x^n_1)} \nonumber \\
& = \frac{\bar {\mu}_\rho(x^n_1)}{\bar {\mu}_\rho(x^n_2)}. \label{eq:mux2tomux1}
\end{align}
Moreover,
\begin{align}
\mu_\rho(\{x^n_2\}) & = \bar {\mu}_\rho(x^n_2) - \bar {\mu}_\rho(x^n_1) \nonumber \\
& = \bar {\mu}_\rho(x^n_2) \left( 1 - \frac{\bar {\mu}_\rho(x^n_1)}{\bar {\mu}_\rho(x^n_2)} \right) \nonumber \\
& = \bar {\mu}_\rho(x^n_2) \left( 1 - \frac{\psi_{\mu}(x^n_2) - x^n_2}{\psi_{\mu}(x^n_1)-x^n_2} \right) \nonumber \\
& = \bar {\mu}_\rho(x^n_2) \frac{\psi_{\mu}(x^n_1)-\psi_{\mu}(x^n_2)}{\psi_{\mu}(x^n_1)-x^n_2}. \label{eq:mequalone:mux1}
\end{align}

First, $x^n_2$ shall be fully embedded before $\mathcal{H}_{n+1}$. Then
        \begin{equation}\label{eq:mequalone:x1fullyembedded}
        \begin{aligned}
        \mu_\rho(\{x^n_2\}) = & \mathbb{P}(\widetilde S_{\tau} = x^n_2 \text{ and } \tau < \mathcal{H}_{n+1})  \\
        = & \mathbb{P}(\widetilde S_{\tau} = x^n_2 \text{ and } \tau < \mathcal{H}_{n}) + \mathbb{P}(\widetilde S_{\tau} = x^n_2 \text{ and } \mathcal{H}_{n} \leq \tau < \mathcal{H}_{n+1})\\
        = & \mathbb{P}(\widetilde S_{\tau} = x^{n-1}_1 \text{ and } \tau < \mathcal{H}_{n}) + \mathbb{P}(\widetilde S_{\tau} = x^n_2 \text{ and } \tau < \mathcal{H}_{n+1}|\tau \geq \mathcal{H}_{n}) \cdot \mathbb{P}(\tau \geq \mathcal{H}_{n})\\
        = & \bar {\mu}_\rho(x^{n-1}_1)\frac{g_\rho(n)-\psi_{\mu}(x^{n-1}_1)}{g_\rho(n)-x^{n-1}_1}+\frac{g_\rho(n+1)-g_\rho(n)}{g_\rho(n+1)-x^n_2}(1-r^n_1)\cdot\bar {\mu}_\rho(x^{n-1}_1)\frac{\psi_{\mu}(x^{n-1}_1)-x^{n-1}_1}{g_\rho(n)-x^{n-1}_1}\\
        = & \bar {\mu}_\rho(x^{n}_2)\left(\frac{g_\rho(n)-\psi_{\mu}(x^{n}_2)}{g_\rho(n)-x^{n}_2}+\frac{g_\rho(n+1)-g_\rho(n)}{g_\rho(n+1)-x^{n}_2}\frac{\psi_{\mu}(x^{n}_2)-x^{n}_2}{g_\rho(n)-x^{n}_2}(1-r^n_1)\right),
        \end{aligned}
        \end{equation}
where the second last equality uses \eqref{eq:stoppingtimedistribution}.
Equating the right hand side of \eqref{eq:mequalone:x1fullyembedded} with the right hand side of \eqref{eq:mequalone:mux1}
and solving for $r^n_1$, we obtain
\begin{align}\label{eq:probabilityrn2}
r^n_1 = \frac{(g_\rho(n)-x^n_2)(g_\rho(n+1)-\psi_{\mu}(x^n_1))}{(\psi_{\mu}(x^n_1)-x^n_2)(g_\rho(n+1) - g_\rho(n))}.
\end{align}
To verify that above $r^n_1$ is indeed a probability, note that $r^n_1$ is a decreasing function of $\psi_{\mu}(x^n_1)$. Recall that $x^n_1 = b_\mu(g_\rho(n+1))$. By definition of $\psi_{\mu}$, $b_\mu$ and the fact that $m_n = 1$,   $\psi_{\mu}(b_\mu(g_\rho(n+1))) \in [g_\rho(n), g_\rho(n+1)]$. Then we have $r^n_1 \in [0,1]$.

Next, $x^n_1$ is not stopped before $\mathcal{H}_{n}$ and shall be partially embedded before $\mathcal{H}_{n+1}$.
We have
\begin{align*}
&\mathbb{P}(\widetilde S_{\tau} = x^n_1 \text{ and } \tau < \mathcal{H}_{n+1}) \\
=& \mathbb{P}(\widetilde S_{\tau} = x^n_1 \text{ and } \tau < \mathcal{H}_{n+1} | \tau \geq \mathcal{H}_{n}) \cdot \mathbb{P}(\tau \geq \mathcal{H}_{n})\\
=& \frac{g_\rho(n+1)-g_\rho(n)}{g_\rho(n+1)-x^n_1} r^n_1 \cdot \bar {\mu}_\rho(x^{n-1}_1) \frac{\psi_{\mu}(x^{n-1}_1)-x^{n-1}_1}{g_\rho(n)-x^{n-1}_1}\\
=& \frac{g_\rho(n+1)-g_\rho(n)}{g_\rho(n+1)-x^n_1} \frac{(g_\rho(n)-x^n_2)(g_\rho(n+1)-\psi_{\mu}(x^n_1))}{(\psi_{\mu}(x^n_1)-x^n_2)(g_\rho(n+1)-g_\rho(n))} \cdot \frac{\psi_{\mu}(x^n_1)-x^n_2}{\psi_{\mu}(x^n_2)-x^n_2}\bar {\mu}_\rho(x^n_1) \frac{\psi_{\mu}(x^n_2)-x^n_2}{g_\rho(n)-x^n_2}\\
=& \bar {\mu}_\rho(x^n_1) \frac{g_\rho(n+1)-\psi_{\mu}(x^n_1)}{g_\rho(n+1)-x^n_1},
\end{align*}
where the second equality uses \eqref{eq:stoppingtimedistribution} and the third equality uses \eqref{eq:mux2tomux1} and \eqref{eq:probabilityrn2}.

Finally, for $x > x^n_1$, note that they are not stopped during $\mathcal{H}_{n} \le \tau < \mathcal{H}_{n+1}$, then
\begin{align*}
\mathbb{P}(\widetilde S_{\tau} = x \text{ and } \tau < \mathcal{H}_{n+1}) = \mathbb{P}(\widetilde S_{\tau} = x \text{ and } \tau < \mathcal{H}_{n}) = 0.
\end{align*}
We conclude that when $m_n = 1$, (\ref{eq:AYlikestoppingtime}) is satisfied for $j = n$.

\noindent{\bf Case 3: $m_n \ge 2$}

In the general case, $x^n_1 > \cdots > x^n_{m_n} > x^n_{m_n+1} = x^{n-1}_1$. According to the stopping rule, once the process $(\widetilde S_t, t \in \bN)$ arrives a new maximum $g_\rho(n)$, set the upper bound to be $g_\rho(n+1)$ and the lower bound $x^n_{1}$. If $x^n_{1}$ is reached first, toss a coin with specified probability of tail/head. If it shows a tail, stop; if it shows a head, update the lower bound to be $x^n_{2}$ and continue. Once $x^n_{2}$ is reached, toss a coin with specified probability of tail/head. This procedure continues until either a new maximum $g_\rho(n+1)$ is reached, or by tossing a coin tail, stop at state $x^n_{k}$ for some $k = 1,2,...,m_n$, or, finally, stop at $x^n_{m_n+1}$ for sure. Then, conditional on $\tau \geq \mathcal{H}_n$,
\begin{align*}
\tau \wedge \mathcal{H}_{n+1} = \begin{cases}
\inf\Big\{t \geq \mathcal{H}_{n}: \widetilde S_t \in \{x^n_{1}, g_\rho(n+1)\}\Big\} & \text{ if } \xi^n_{1} = 1\\
\inf\Big\{t \geq \mathcal{H}_{n}: \widetilde S_t \in \{x^n_{2},g_\rho(n+1)\}\Big\} & \text{ if } \xi^n_{1} = 0 \text{ and }\xi^n_{2} = 1\\
\vdots\\
\inf\Big\{t \geq \mathcal{H}_{n}: \widetilde S_t \in \{x^n_{m_n},g_\rho(n+1)\}\Big\} & \text{ if } \xi^n_{1} = ... = \xi^n_{m_n-1} = 0 \text{ and }\xi^n_{m_n} = 1\\
\inf\Big\{t \geq \mathcal{H}_{n}: \widetilde S_t \in \{x^n_{m_n+1},g_\rho(n+1)\}\Big\} & \text{ if } \xi^n_{1} = ... = \xi^n_{m_n} = 0,
\end{cases}
\end{align*}
where $\mathbb{P}(\xi^n_k = 1) = r^n_k$, the probabilities of tossing a tail at state $x^n_k$, $k = 1, 2,...,m_n$, to be determined.

Analogous to \eqref{eq:mux2tomux1} and \eqref{eq:mequalone:mux1}, we have
\begin{equation}\label{eq:muxktomux1}
\begin{aligned}
& \bar {\mu}_\rho(x^n_k) = \bar{\mu}_\rho(x^n_{m_n+1}) \left[ \prod_{l = k}^{m_n}   \frac {\bar{\mu}_\rho(x^{n}_{l})}{\bar {\mu}_\rho(x^n_{l+1})} \right] = \bar{\mu}_\rho(x^{n-1}_1) \left[ \prod_{l = k}^{m_n} \frac{\psi_{\mu}(x^n_{l+1}) - x^n_{l+1}}{\psi_{\mu}(x^n_{l}) - x^n_{l+1}} \right].  \\
& \mu_\rho(\{x^n_k\})  = \bar {\mu}_\rho(x^n_k) \frac{\psi_{\mu}(x^n_{k-1})-\psi_{\mu}(x^n_k)}{\psi_{\mu}(x^n_{k-1})-x^n_k} = \bar{\mu}_\rho(x^{n-1}_1) \left[ \prod_{l = k}^{m_n} \frac{\psi_{\mu}(x^n_{l+1}) - x^n_{l+1}}{\psi_{\mu}(x^n_{l}) - x^n_{l+1}} \right] \frac{\psi_{\mu}(x^n_{k-1})-\psi_{\mu}(x^n_k)}{\psi_{\mu}(x^n_{k-1})-x^n_k} .
\end{aligned}
\end{equation}

First, $x^n_2, x^n_3,...,x^n_{m_n+1}$ should be fully embedded before $\mathcal{H}_{n+1}$. For  $x^n_2,...,x^n_{m_n}$, since they are not hit before $\mathcal{H}_{n}$, they should be fully embedded between $\mathcal{H}_{n}$ and $\mathcal{H}_{n+1}$. Then
        \begin{align*}
        \mu_\rho(\{x^n_k\}) = & \mathbb{P}(\widetilde S_{\tau} = x^n_k \text{ and } \mathcal{H}_n \leq \tau < \mathcal{H}_{n+1}) \\
        = & \mathbb{P}(\widetilde S_{\tau} = x^n_k \text{ and } \tau < \mathcal{H}_{n+1} | \tau \geq \mathcal{H}_n) \cdot \mathbb{P}(\tau \geq \mathcal{H}_n)\\
        = & \frac{g_\rho(n+1)-g_\rho(n)}{g_\rho(n+1)-x^n_k}(1-r^n_{1}) \cdots (1-r^n_{k-1})r^n_k \cdot \bar{\mu}_\rho(x^{n-1}_1)\frac{\psi_{\mu}(x^{n-1}_1)-x^{n-1}_1}{g_\rho(n)-x^{n-1}_1}, \\
        & k = 2,...,m_n.
        \end{align*}
where the last equality uses \eqref{eq:stoppingtimedistribution}.
Then we have $m_n - 1$ equations:
\begin{align*}
(1-r^n_{1}) \cdots (1-r^n_{k-1})r^n_k =  \left[ \prod_{l = k}^{m_n} \frac{\psi_{\mu}(x^n_{l+1}) - x^n_{l+1}}{\psi_{\mu}(x^n_{l}) - x^n_{l+1}} \right]  \frac{\psi_{\mu}(x^n_{k-1})-\psi_{\mu}(x^n_k)}{\psi_{\mu}(x^n_{k-1} )-x^n_k} \\
\cdot \frac{g_\rho(n)-x^{n}_{m_n+1}}{\psi_{\mu}(x^{n}_{m_n+1})-x^{n}_{m_n+1}} \frac{g_\rho(n+1)-x^n_k}{g_\rho(n+1)-g_\rho(n)}, \\
k = 2,...,m_n.
\end{align*}
The last equation is derived from embedding $x^n_{m_n+1}$. Note that $x^n_{m_n+1}$ is partially embedded before $\mathcal{H}_{n}$ and should be fully embedded before $\mathcal{H}_{n+1}$. Then
        \begin{align*}
        \mu_\rho(\{x^n_{m_n+1}\}) = & \mathbb{P}(\widetilde S_{\tau} = x^n_{m_n+1} \text{ and } \tau < \mathcal{H}_n) + \mathbb{P}(\widetilde S_{\tau} = x^n_{m_n+1} \text{ and } \mathcal{H}_n \leq \tau < \mathcal{H}_{n+1}) \\
        = & \mathbb{P}(\widetilde S_{\tau} = x^n_{m_n+1} \text{ and } \tau < \mathcal{H}_n) + \mathbb{P}(\widetilde S_{\tau} = x^n_{m_n+1} \text{ and } \tau < \mathcal{H}_{n+1}) | \tau \geq \mathcal{H}_n) \cdot \mathbb{P}(\tau \geq \mathcal{H}_n)\\
        = & \bar {\mu}_\rho(x^n_{m_n+1}) \frac{g_\rho(n)-\psi_{\mu}(x^n_{m_n+1})}{g_\rho(n)-x^n_{m_n+1}}  + \frac{g_\rho(n+1)-g_\rho(n)}{g_\rho(n+1)-x^n_{m_n+1}} (1-r^n_1) \cdots (1-r^n_{m_n}) \\
        & \quad \cdot \bar{\mu}_\rho(x^n_{m_n+1})\frac{\psi_{\mu}(x^n_{m_n+1})-x^n_{m_n+1}}{g_\rho(n)-x^n_{m_n+1}},
        \end{align*}
which leads to
\begin{align*}
(1-r^n_{1}) \cdots (1-r^n_{m_n}) = \frac{\psi_{\mu}(x^n_{m_n}) - g_\rho(n)}{\psi_{\mu}(x^n_{m_n})-x^n_{m_n+1}} \frac{g_\rho(n+1)-x^n_{m_n+1}}{g_\rho(n+1)-g_\rho(n)}.
\end{align*}
This yields totally $m_n$ equations to solve $m_n$ unknowns, $r^n_k$, $k = 1,...,m_n$.
We obtain
\begin{align*}
\begin{cases}
& r^n_1 = \frac{g_\rho(n+1) - \psi_{\mu}(x^n_{1})}{g_\rho(n+1)-g_\rho(n)}  \left[ \prod_{l = 1}^{m_n} \frac{\psi_{\mu}(x^n_{l+1}) - x^n_{l+1}}{\psi_{\mu}(x^n_{l}) - x^n_{l+1}} \right] \frac{g_\rho(n)-x^n_{m_n+1}}{\psi_{\mu}(x^n_{m_n+1})-x^n_{m_n+1}} , \\
& r^n_{k} = \frac{\left[ \prod_{l = k}^{m_n} \frac{\psi_{\mu}(x^n_{l+1}) - x^n_{l+1}}{\psi_{\mu}(x^n_{l}) - x^n_{l+1}} \right]  \frac{\psi_{\mu}(x^n_{k-1})-\psi_{\mu}(x^n_{k})}{\psi_{\mu}(x^n_{k-1} )-x^n_{k}}\frac{g_\rho(n)-x^n_{m_n+1}}{\psi_{\mu}(x^n_{m_n+1})-x^n_{m_n+1}} \frac{g_\rho(n+1)-x^n_{k}}{g_\rho(n+1)-g_\rho(n)}}{\sum_{h = k}^{m_n} \left[ \prod_{l = h}^{m_n} \frac{\psi_{\mu}(x^n_{l+1}) - x^n_{l+1}}{\psi_{\mu}(x^n_{l}) - x^n_{l+1}} \right]  \frac{\psi_{\mu}(x^n_{h-1})-\psi_{\mu}(x^n_h)}{\psi_{\mu}(x^n_{h-1} )-x^n_h}\frac{g_\rho(n)-x^n_{m_n+1}}{\psi_{\mu}(x^n_{m_n+1})-x^n_{m_n+1}} \frac{g_\rho(n+1)-x^n_h}{g_\rho(n+1)-g_\rho(n)} + \frac{\psi_{\mu}(x^n_{m_n}) - g_\rho(n)}{\psi_{\mu}(x^n_{m_n})-x^n_{m_n+1}} \frac{g_\rho(n+1)-x^n_{m_n+1}}{g_\rho(n+1)-g_\rho(n)}}, \\
& \qquad k = 2,...,m_n.
\end{cases}
\end{align*}
Moreover, $r^n_k \in [0,1]$, $k = 1,...,m_n$.

Next, $x^{n}_1$ is not stopped before $\mathcal{H}_{n}$ and shall be partially embedded before $\mathcal{H}_{n+1}$.
We have
\begin{align*}
&\mathbb{P}(\widetilde S_{\tau} = x^n_{1} \text{ and } \tau < \mathcal{H}_{n+1}) \nonumber \\
=& \mathbb{P}(\widetilde S_{\tau} = x^n_{1} \text{ and } \tau < \mathcal{H}_{n+1} | \tau \geq \mathcal{H}_{n}) \cdot \mathbb{P}(\tau \geq \mathcal{H}_{n}) \nonumber \\
=& \frac{g_\rho(n+1)-g_\rho(n)}{g_\rho(n+1)-x^n_{1}} r^n_{1} \cdot \bar {\mu}_\rho(x^{n-1}_1) \frac{\psi_{\mu}(x^{n-1}_1)-x^{n-1}_1}{g_\rho(n)-x^{n-1}_1} \nonumber \\
=& \frac{g_\rho(n+1)-g_\rho(n)}{g_\rho(n+1)-x^n_{1}} r^n_{1}  \cdot  \left[ \prod_{l = 1}^{m_n} \frac{\psi_{\mu}(x^n_{l}) - x^n_{l+1}}{\psi_{\mu}(x^n_{l+1}) - x^n_{l+1}} \right] \bar {\mu}_\rho(x^n_{1}) \frac{\psi_{\mu}(x^n_{m_n+1})-x^n_{m_n+1}}{g_\rho(n)-x^n_{m_n+1}} \\
=& \bar {\mu}_\rho(x^n_{1}) \frac{g_\rho(n+1) - \psi_{\mu}(x^n_{1})}{g_\rho(n+1)-x^n_{1}},
\end{align*}
where the second last equality uses \eqref{eq:muxktomux1} with $k = m_n+1$, and the last equality is by plugging $r^n_{1}$ into the equation.

Finally, for $x > x^n_{1}$, note that they are not stopped during $\mathcal{H}_{n} \le \tau < \mathcal{H}_{n+1}$, then
\begin{align*}
\mathbb{P}(\widetilde S_{\tau} = x \text{ and } \tau < \mathcal{H}_{n+1}) = \mathbb{P}(\widetilde S_{\tau} = x \text{ and } \tau < \mathcal{H}_{n}) = 0.
\end{align*}
We conclude that when $m_n \ge 2$, (\ref{eq:AYlikestoppingtime}) is satisfied for $j = n$.
\halmos

\pfof{Theorem \ref{thm:UnfairGames:UI}}
Suppose the equality of the last constraint in \eqref{prob:UnfairGames:infinitedim} does not hold under ${\bf z}^*$:
\begin{align*}
\frac{1}{1-p} \sum_{n=1}^\infty \rho^{n} z_{+,n}^* - \frac{1}{p} \sum_{n=1}^\infty  \rho^{-n} z_{-,n}^* < 0.
\end{align*}
Then we could slightly increase $z_{+,n}^*$ for some $n$ without violating the other constraints of \eqref{prob:UnfairGames:infinitedim}. Consequently, the objective value is strictly increased, which violates the optimality of ${\bf z}^*_{+,-}$. Hence, ${\bf z}^*_{+,-}$ should make the equality holds. This shows that the corresponding probability measure $\mu^*$ belong to $\mathcal{\widehat M}^\rho$ as well as $\mathcal{M}^\rho$. Then there exists a randomized stopping time $\tau^* \in {\cal T}$ such that $S_{\tau^*} \sim \mu^*$ and $(S_{\tau^* \wedge t}, t \in \mathbb{N})$ is uniformly integrable.
Moreover, according to Lemma \ref{le:uniformintegrable},
\begin{align*}
\expect[\tau^*] = \frac{1}{2p-1} \expect[S_{\tau^*}] = \frac{1}{2p-1} \left( \sum_{n=1}^\infty z_{+,n}^* - \sum_{n=1}^\infty z_{-,n}^* \right).
\end{align*}
\halmos

\pfof{Proposition \ref{prop:wellposedness}}
For any feasible solution ${\bf z_{+,-}} = (z_{+,1},z_{+,2},...;z_{-,1},z_{-,2},...)$ to \eqref{prob:UnfairGames:infinitedim}, we have
\begin{align*}
\frac{1}{1-p} \sum_{n=1}^\infty \rho^{n} z_{+,n} \leq \frac{1}{p} \sum_{n=1}^\infty \rho^{-n} z_{-,n} \leq \frac{1}{p} \sum_{n=1}^\infty \rho^{-n} = \frac{1}{1-2p}.
\end{align*}
Then
\begin{align*}
z_{+,n} \left(\frac{1}{1-p} \sum_{j=1}^n \rho^{j} \right) & = \frac{1}{1-p} \left( \sum_{j=1}^n \rho^{j} z_{+,n} \right) \\
& \leq \frac{1}{1-p} \left( \sum_{j=1}^n \rho^{j} z_{+,j} \right) \leq \frac{1}{1-p} \left( \sum_{n=1}^\infty \rho^{j} z_{+,j} \right) \leq \frac{1}{1-2p} \; ,
\end{align*}
which implies that
\begin{align*}
z_{+,n} \leq \frac{1}{1-2p} \frac{1-p}{ \sum_{j=1}^n \rho^{j} } =  \frac{1}{\rho^n -1 }  \;.
\end{align*}

On the other hand, by Tonelli's theorem the first sum of infinite series in $v({\bf z_{+,-}})$ could be written as
\begin{align*}
& \sum_{n=1}^\infty u_+(n) \Big(w_+(z_{+,n})-w_+(z_{+,n+1})\Big) \\
= & \sum_{n=1}^\infty \sum_{j = 1}^n \Big( u_+(j) - u_+(j-1) \Big)  \Big(w_+(z_{+,n})-w_+(z_{+,n+1})\Big) \\
= & \sum_{j = 1}^\infty \sum_{n=j}^\infty \Big( u_+(j) - u_+(j-1) \Big)  \Big(w_+(z_{+,n})-w_+(z_{+,n+1})\Big) \\
= & \sum_{j = 1}^\infty \Big( u_+(j) - u_+(j-1) \Big) w_+(z_{+,j}) \;.
\end{align*}
Then
\begin{align*}
v({\bf z_{+,-}}) \leq \sum_{n = 1}^\infty \Big(u_+(n)-u_+(n-1)\Big) w_+(z_{+,n}) \le \sum_{n = 1}^\infty \Big(u_+(n)-u_+(n-1)\Big) w_+\left( \frac{1}{\rho^n-1} \right).
\end{align*}
Since $|u_+(n)-u_+(n-1)|$ is bounded for $n \in \bN$, there exists $\bar M > 0$ such that
\begin{align*}
\sum_{n = 1}^\infty \Big(u_+(n)-u_+(n-1)\Big) w_+\left( \frac{1}{\rho^n-1} \right) \le \bar M \sum_{n = 1}^\infty w_+\left( \frac{1}{\rho^n-1} \right),
\end{align*}
which is finite due to \eqref{eq:UnfairGames:wellposedness}.
Therefore, $v({\bf z_{+,-}})$ is upper-bounded for any ${\bf z_{+,-}}$ and problem (\ref{prob:UnfairGames:infinitedim}) has finite objective value. \halmos

\pfof{Proposition \ref{prop:UnfairGames:GainPart}}
Introducing a Lagrange multiplier $\mu$ into the objective function, we have
\begin{align*}
v_+({\bf z};s) - \mu (\sum_{n=1}^\infty \rho^{n} {z_n} - s)=\sum_{n=1}^\infty \left( \Big((n+1)^{\alpha_+}-n^{\alpha_+}\Big)z_n^{\delta_+} - \mu \rho^n z_n \right) + \mu s .
\end{align*}
Maximizing above over $z_n$ individually with constraints, we have
\begin{align*}
z_n^*(\mu)=\left( \frac{\delta_+}{\mu} \frac{1}{\rho^n} \Big((n+1)^{\alpha_+}-n^{\alpha_+}\Big)\right)^{\frac{1}{1-\delta_+}} \wedge1, \quad n=1,2,...
\end{align*}
Note that $z_n^*(\mu)$ is decreasing in $n$, then $z_1^*(\mu) \ge z_2^*(\mu)... \ge z_n^*(\mu) \ge ...$.
The Lagrange multiplier $\mu = \mu(s)$ is such that
\begin{align*}
\sum_{n=1}^\infty \left( \frac{\delta_+}{\mu(s)} \frac{1}{\rho^n} \Big((n+1)^{\alpha_+}-n^{\alpha_+}\Big) \right)^{\frac{1}{1-\delta_+}} \wedge1=s.
\end{align*}
The optimal value of (\ref{prob:UnfairGames:GainPart}), as a function of $s$, denoted by $v_{+}^*(s)$, is
\begin{align*}
v_{+}^*(s) = v_+({\bf z}^*;s) = \sum_{n=1}^\infty \left( \left(\frac{\delta_+}{\mu(s)}\right)^{\delta_+} \frac{1}{\rho^{n \delta_+}} \Big((n+1)^{\alpha_+}-n^{\alpha_+}\Big)\right)^\frac{1}{1-\delta_+}\wedge\Big((n+1)^{\alpha_+}-n^{\alpha_+}\Big).
\end{align*}
If $0 \le s < A_1 \left(\frac{1}{\rho}(2^{\alpha_+}-1) \right)^\frac{1}{\delta_+-1}$,
then $\mu = \mu(s)$ is such that $1 > z_1^*(\mu) > z_2^*(\mu) > ...$. Then
\begin{align*}
& \sum_{n=1}^\infty \left( \frac{\delta_+}{\mu(s)} \frac{1}{\rho^n} \Big((n+1)^{\alpha_+}-n^{\alpha_+}\Big) \right)^{\frac{1}{1-\delta_+}} =s, \\
\Rightarrow & \left( \frac{\delta_+}{\mu(s)} \right)^{\frac{1}{1-\delta_+}} = \frac{s}{ \sum_{n=1}^\infty \left( \frac{1}{\rho^n} \Big((n+1)^{\alpha_+}-n^{\alpha_+}\Big) \right)^{\frac{1}{1-\delta_+}}} = s A_1^{-1}.
\end{align*}
Hence,
\begin{align*}
& z_n^* = \left( \frac{\delta_+}{\mu(s)} \frac{1}{\rho^n} \Big((n+1)^{\alpha_+}-n^{\alpha_+}\Big)\right)^{\frac{1}{1-\delta_+}}  = s A_1^{-1} \left( \frac{1}{\rho^n} \Big((n+1)^{\alpha_+}-n^{\alpha_+}\Big)\right)^{\frac{1}{1-\delta_+}} , \quad n=1,2,... \\
& v_{+}^*(s) = \sum_{n=1}^\infty \left( \left(\frac{\delta_+}{\mu(s)}\right)^{\delta_+} \frac{1}{\rho^{n \delta_+}} \Big((n+1)^{\alpha_+}-n^{\alpha_+}\Big)\right)^\frac{1}{1-\delta_+} = (s A_1^{-1})^{\delta_+} W_1.
\end{align*}
If $j-1 + A_{j} \left( \frac{1}{\rho^j} ( (j+1)^{\alpha_+}-{j}^{\alpha_+}) \right)^\frac{1}{\delta_+-1} \le s < j + A_{j+1} \left( \frac{1}{\rho^{j+1}} ({(j+2)}^{\alpha_+}-{(j+1)}^{\alpha_+}) \right)^\frac{1}{\delta_+-1}$ for some $j \ge 1$,
then $\mu = \mu(s)$ is such that $z_1^*(\mu) = ... = z_j^*(\mu) = 1$, $1 > z_{j+1}^*(\mu) > ...$. Then
\begin{align*}
& \sum_{n = 1}^j 1 + \sum_{n=j+1}^\infty \left( \frac{\delta_+}{\mu(s)} \frac{1}{\rho^n} \Big((n+1)^{\alpha_+}-n^{\alpha_+}\Big) \right)^{\frac{1}{1-\delta_+}} =s, \\
\Rightarrow & \left( \frac{\delta_+}{\mu(s)} \right)^{\frac{1}{1-\delta_+}} = \frac{s-j}{ \sum_{n=j+1}^\infty \left( \frac{1}{\rho^n} \Big((n+1)^{\alpha_+}-n^{\alpha_+}\Big) \right)^{\frac{1}{1-\delta_+}}} = (s-j) A_{j+1}^{-1}.
\end{align*}
Hence,
\begin{align*}
& z_n^* =
\begin{cases}
1 &, n = 1,...,j \\
\left( \frac{\delta_+}{\mu(s)} \frac{1}{\rho^n} \Big((n+1)^{\alpha_+}-n^{\alpha_+}\Big)\right)^{\frac{1}{1-\delta_+}}  = (s-j)A_{j+1}^{-1} \left( \frac{1}{\rho^n} \Big((n+1)^{\alpha_+}-n^{\alpha_+}\Big)\right)^{\frac{1}{1-\delta_+}} & , n=j+1,...
\end{cases}
\\
& v_{+}^*(s)  = \sum_{n = 1}^j \Big((n+1)^{\alpha_+}-n^{\alpha_+}\Big) + \sum_{n=j+1}^\infty \left( \left(\frac{\delta_+}{\mu(s)}\right)^{\delta_+} \frac{1}{\rho^{n \delta_+}} \Big((n+1)^{\alpha_+}-n^{\alpha_+}\Big)\right)^\frac{1}{1-\delta_+} \\
& = (j+1)^{\alpha_+}-1 + (s-j)^{\delta_+} {A_{j+1}}^{-\delta_+} W_{j+1}.
\end{align*}
\halmos

\pfof{Proposition \ref{prop:UnfairGames:GainPart:valuefunction}}
Let $m$ be the largest integer such that $\frac{\rho(\rho^{m}-1)}{\rho - 1} \leq s$. Then
\begin{align*}
\frac{\log( (\rho-1)s +\rho)} {\log \rho} - 2 < m \leq \frac{\log( (\rho-1)s +\rho)} {\log \rho} - 1.
\end{align*}
Moreover,
\begin{align*}
0 > s - \frac{\rho(\rho^{m+1}-1)}{\rho - 1} = s - \frac{\rho(\rho^{m}-1)}{\rho - 1} - \rho^{m+1}, \\
\Rightarrow \rho^{m+1} > s - \frac{\rho(\rho^{m}-1)}{\rho - 1}.
\end{align*}
Consider the following feasible solution ${\bf z}$ to \eqref{prob:UnfairGames:GainPart}:
\begin{align*}
z_1 = z_2 =...=z_{m} = 1, z_{m+1} = \frac{1}{\rho^{m+1}} \left(s - \frac{\rho(\rho^{m}-1)}{\rho - 1} \right), z_{m+2} =...= 0.
\end{align*}
Hence, there exists some positive constant $c_{+}$ such that
\begin{align*}
v_{+}^*(s) \ge v_+({\bf z};s) \ge (m+1)^{\alpha_+}-1 &> \left(\frac{\log( (\rho-1)  s +  \rho)} {\log \rho} - 1 \right)^{\alpha_+}-1 \\
&>  \left(\frac{\log (\rho-1) + \log s } {\log \rho} - 1 \right)^{\alpha_+}-1 \\
&> c_{+} ({\log s})^{\alpha_+},
\end{align*}
for sufficiently large values of $s$.

To show the opposite bound, we rewrite the problem \eqref{prob:UnfairGames:GainPart} in terms of the quantile formulation.
For any feasible solution ${\bf z} = (z_1,z_2,...)$ to \eqref{prob:UnfairGames:GainPart}, let $F: [0,+\infty) \mapsto [0,1]$ be right-continuous step function such that $F(n) = 1 - z_{n}$ for $n \in \bN$, where $z_0 : =1$.
Then there exists a corresponding inverse function of $F$, denoted by $G: [0,1] \mapsto \bN$.
Let $G(x) := \inf\{y \in \bN : F(y) \ge x\}$, $x \ge 0$. Hence, $G(x)$ is a left-continuous step function.
The objective function of \eqref{prob:UnfairGames:GainPart} is then written in terms of the quantile function:
\begin{align*}
v_{+,Q}(G(\cdot)) := v_+({\bf z}) & = \sum_{n = 1}^\infty \Big(u_+(n+1) - u_+(n) \Big) w_+(z_n) = \sum_{n = 1}^\infty u_+(n) \Big (w_+(z_{n-1}) - w_+(z_{n}) \Big) - 1\\
& = \int_0^{+\infty} u_+(y) d w_+(1 - F(y)) - 1 \\
& = \int_0^{1} u_+(G(x)) d w_+(1 - x) - 1 = \int_0^1 u_+(G(x))w_+^{'}(1-x) dx - 1.
\end{align*}
Denote by $\mathcal{G_+}$ the set of left-continuous step quantile functions taking value in non-negative integers.
Then Problem \eqref{prob:UnfairGames:GainPart} is equivalent to
\begin{equation}\label{prob:UnfairGames:GainPart:quantile}
\begin{array}{cl}
\underset{G(\cdot) \in \mathcal{G_+}}{\text{sup }} &v_{+,Q}(G(\cdot);s) \\
\text{subject to  } &\int_0^1 \rho^{G(x)} dx \leq (\rho-1) s + \rho. \tag{Q-GP}
\end{array}
\end{equation}
Denote ${\cal A}_{+}(s) := \{G(\cdot) \in \mathcal{G_+} : \int_0^1 \rho^{G(x)} dx \leq (\rho-1) s + \rho \}$.
Let $s$ be sufficiently large, and $G_0(\cdot) \in {\cal A}_{+}(s)$ is a feasible solution to \eqref{prob:UnfairGames:GainPart:quantile}.
Let $n := \lfloor \log s / \log \rho \rfloor > 1$. Then $n + 1 > \log s / \log \rho \ge n$.
Let $\widetilde{G}_0(x) := \frac{G_0(x)}{\log s / \log \rho}$. Note that $\widetilde{G}_0(\cdot)$ may not be feasible as it can be non integer-valued. Consider the integer-valued quantile $\lfloor \widetilde{G}_0(\cdot) \rfloor $. Then $\lfloor \widetilde{G}_0(x) \rfloor \leq \widetilde{G}_0(x) < \lfloor \widetilde{G}_0(x) \rfloor +1$.
As $n$ sufficiently large,
\begin{align*}
(\rho-1) + \rho &\geq (\rho^{n+2} )^{\frac{1}{n}} > \left( (\rho - 1) \rho^{n+1} + \rho \right)^{\frac{1}{n}} > \left( (\rho-1) s + \rho \right)^{\frac{1}{n}} \\
&\geq \left( \int_0^1 \rho^{G_0(x)} dx \right)^{\frac{1}{n}} \geq \left( \int_0^1 \rho^{G_0(x)} dx \right)^{\frac{1}{\log s / \log \rho}} \\
&= \left( \int_0^1 \left( \rho^{\widetilde G_0(x)} \right)^{\log s / \log \rho} dx \right)^{\frac{1}{\log s / \log \rho}} \geq \int_0^1 \rho^{\widetilde G_0(x)} dx \geq \int_0^1 \rho^{\lfloor \widetilde{G}_0(x) \rfloor} dx.
\end{align*}
Hence, $\lfloor \widetilde{G}_0(\cdot) \rfloor  \in {\cal A}_{+}(1)$.
Then
\begin{align*}
v_{+,Q}(G_0(\cdot)) &= \int_0^1 u_+(G_0(x))w_+^{'}(1-x) dx - 1 \\
& =  (\log s / \log \rho)^{\alpha_+} \int_0^1 u_+ \left(\frac{G_0(x)}{\log s / \log \rho} \right)w_+^{'}(1-x) dx -1 \\
& < (\log s / \log \rho)^{\alpha_+} \int_0^1 u_+(\lfloor \widetilde{G}_0(x) \rfloor  + 1)w_+^{'}(1-x) dx \\
&\leq (\log s / \log \rho)^{\alpha_+} \Big( \int_0^1 u_+(\lfloor \widetilde{G}_0(x) \rfloor )w_+^{'}(1-x) dx + 1 \Big)\\
&\leq  (\log s / \log \rho)^{\alpha_+} \Big( v_{+}^{*}(1) + 2 \Big),
\end{align*}
where the second last inequality follows from the concavity of $u_+(\cdot)$. Since $G_0(\cdot)$ is arbitrary, there exists a positive constant $C_{+}$ such that $v_{+}^*(s) < C_{+}(\log s)^{\alpha_+}$.
\halmos

\pfof{Proposition \ref{prop:UnfairGames:infinites01}}
Given a solution ${\bf z} = (z_1,z_2,...) \in \mathcal{Z}_A(s)$, let
\begin{align*}
t_k := \frac{(z_k-z_{k+1})\phi(1-\phi^k)}{s(1-\phi)}, \quad k \ge 1.
\end{align*}
Then $t_k \geq 0$ and
\begin{align*}
\sum_{k=1}^\infty t_k = \sum_{k=1}^{\infty} \frac{(z_k-z_{k+1})\phi(1-\phi^k)}{s(1-\phi)}= \frac{\phi}{s(1-\phi)} \sum_{k=1}^\infty z_k(\phi^{k-1}-\phi^k) =\frac{1}{s}\sum_{k = 1}^\infty \phi^k z_k=1.
\end{align*}
Hence, $t_k \in [0,1]$, $k \ge 1$. On the other hand, for $n \ge 1$,
\begin{align*}
\sum_{k=1}^\infty t_k z^{0,k}_n = \sum_{k=1}^{\infty} \frac{(z_k-z_{k+1})\phi(1-\phi^k)}{s(1-\phi)}z^{0,k}_n = \sum_{k=n}^{\infty} \frac{(z_k-z_{k+1})\phi(1-\phi^k)}{s(1-\phi)}\frac{s(1-\phi)}{\phi(1-\phi^k)} = z_n,
\end{align*}
which shows that ${\bf z}$ is a convex combination of ${\bf z}^{0,k} \in \mathcal{Z}_-^0(s)$.
Then
\begin{align*}
v_-({\bf z};s) & = \sum_{n=1}^\infty \Big((n+1)^{\alpha_-}-n^{\alpha_-}\Big)z_n^{\delta_-} = \sum_{n=1}^\infty \Big((n+1)^{\alpha_-}-n^{\alpha_-}\Big) \left(\sum_{k=1}^\infty t_k z^{0,k}_n\right)^{\delta_-} \\
& \ge  \sum_{n=1}^\infty \Big((n+1)^{\alpha_-}-n^{\alpha_-}\Big) \left(\sum_{k=1}^\infty t_k (z^{0,k}_n)^{\delta_-}\right) = \sum_{k=1}^\infty t_k v_-({\bf z}^{0,k};s),
\end{align*}
with the equality hold if and only if ${\bf z} = {\bf z}^{0,k}$ for some $k \ge 1$. Hence, ${\bf z}^* \in \mathcal{Z}_-^0(s)$.
\halmos

\pfof{Proposition \ref{prop:UnfairGames:infinitesmminus1m}}
Suppose $\phi (1-\phi^{m-1})/(1-\phi) \le s < \phi (1-\phi^{m})/(1-\phi)$ for $m = h$.
Without loss of generality, suppose any feasible solution ${\bf z} \in\mathcal{Z}_-(s)$ is a convex combination of ${\bf z}^{j,k}$, $0 \le j \le m-1$ and $k \ge m$, for $m = 1,...,h-1$, where $h \ge 2$.

For any feasible solution ${\bf z} = (z_1,z_2,...) \in \mathcal{Z}_-(s)$, let
\begin{align*}
t^0_k = \frac{(z_k-z_{k+1})\phi(1-\phi^k)}{(s-\sum_{n=1}^{m-1}{\phi^n(z_n-z_m)})(1-\phi)}, \quad k \ge m.
\end{align*}
Then $t^0_k \geq 0$ and
\begin{align*}
\sum_{k=m}^\infty t^0_k &= \sum_{k=m}^{\infty} \frac{(z_k-z_{k+1})\phi(1-\phi^k)}{(s-\sum_{n=1}^{m-1}{\phi^n(z_n-z_m)})(1-\phi)} \\
& = \frac{\phi}{(s-\sum_{n=1}^{m-1}{\phi^n(z_n-z_m)})(1-\phi)} \left( \sum_{k=m}^\infty z_k(\phi^{k-1}-\phi^k) + z_m(1-\phi^{m-1}) \right) \\
& = \frac{\phi ((\phi^{-1}-1)(s-\sum_{n=1}^{m-1}{\phi^n z_n }) + z_m(1-\phi^{m-1})}{(s-\sum_{n=1}^{m-1}{\phi^n(z_n-z_m)})(1-\phi)} 
=1.
\end{align*}
Hence, $t^0_k \in [0,1]$, $k \ge m$.
Let
\begin{align*}
{\bf z}^{0}_m : & = \sum_{k=m}^\infty t^0_k {\bf z}^{0,k} = \sum_{k=m}^{\infty} \frac{(z_k-z_{k+1})\phi(1-\phi^k)}{(s-\sum_{n=1}^{m-1}{\phi^n(z_n-z_m)})(1-\phi)} (\underbrace{\frac{s(1-\phi)}{\phi(1-\phi^k)},...,\frac{s(1-\phi)}{\phi(1-\phi^k)}}_{k},0,...)\\
& = \frac{s}{s-\sum_{n=1}^{m-1}{\phi^n(z_n-z_m)}}(\underbrace{ \sum_{k=m}^{\infty} (z_k-z_{k+1}),...\sum_{k=m}^{\infty} (z_k-z_{k+1})}_{m},\sum_{k=m+1}^{\infty} (z_k-z_{k+1}),...) \\
& = (\underbrace{\frac{s}{s-\sum_{n=1}^{m-1}{\phi^n(z_n-z_m)}} z_m,...\frac{s}{s-\sum_{n=1}^{m-1}{\phi^n(z_n-z_m)}} z_m}_{m},\frac{s}{s-\sum_{n=1}^{m-1}{\phi^n(z_n-z_m)}}z_{m+1},...).
\end{align*}
On the other hand, let
\begin{align*}
{\bf z}^{1}_{m-1} := (\underbrace {1,...1}_{m-1}, \frac{s(1-\phi)-{\phi (1-\phi^{m-1})}}{(s-\sum_{n=1}^{m-1}\phi^n z_n)(1-\phi)} z_m, \frac{s(1-\phi)-{\phi (1-\phi^{m-1})} }{(s-\sum_{n=1}^{m-1}\phi^n z_n)(1-\phi)} z_{m+1},...).
\end{align*}
Then ${\bf z}^1_{m-1} \in \mathcal{Z}_-(s)$.
Let ${\bf \hat z}^{1}_{m-1}$ be a sub-vector of ${\bf z}^{1}_{m-1}$ with the first $m-1$ components deleted:
\begin{align*}
{\bf \hat z}^{1}_{m-1} := (\frac{s(1-\phi)-{\phi (1-\phi^{m-1})}}{(s-\sum_{n=1}^{m-1}\phi^n z_n)(1-\phi)} z_m, \frac{s(1-\phi)-{\phi (1-\phi^{m-1})} }{(s-\sum_{n=1}^{m-1}\phi^n z_n)(1-\phi)} z_{m+1},...).
\end{align*}
Note that the geometric summation of components of ${\bf \hat z}^{1}_{m-1}$ equal to
\begin{align*}
\frac{s(1-\phi)-{\phi (1-\phi^{m-1})}}{(s-\sum_{n=1}^{m-1}\phi^n z_n)(1-\phi)} \sum_{n = 1}^\infty \phi^n z_{m+n-1} & = \frac{s(1-\phi)-{\phi (1-\phi^{m-1})}}{(s-\sum_{n=1}^{m-1}\phi^n z_n)(1-\phi)}   \phi^{1-m} \sum_{n = m}^\infty \phi^n z_{n} \\
& = \frac{s(1-\phi)-{\phi (1-\phi^{m-1})}}{(1-\phi)\phi^{m-1}} < \phi.
\end{align*}
Then ${\bf \hat z}^{1}_{m-1} \in \mathcal{Z}_- \left(\frac{s(1-\phi)-{\phi (1-\phi^{m-1})}}{(1-\phi)\phi^{m-1}}\right)$.
By Proposition \ref{prop:UnfairGames:infinites01}, ${\bf \hat z}^{1}_{m-1}$ is a convex combination of ${\bf \hat z}^{0,k'}$, $k' \ge 1$, where
\begin{align*}
{\bf \hat z}^{0,k'} = (\underbrace {\frac {s(1-\phi)-{\phi (1-\phi^{m-1})}}{{\phi^{m} (1-\phi^{k'})}},... \frac {s(1-\phi)-{\phi (1-\phi^{m-1})}}{{\phi^{m} (1-\phi^{k'})}}}_{k'},0...).
\end{align*}
Equivalently, ${\bf z}^{1}_{m-1}$ is a convex combination of ${\bf z}^{m-1,k} \in \mathcal{Z}_-^{m-1}(s)$, $k \ge m$, where
\begin{align*}
{\bf z}^{m-1,k} = (\underbrace {1,1,...1}_{m-1}, \underbrace {\frac {s(1-\phi)-{\phi (1-\phi^{m-1})}}{{\phi^{m} (1-\phi^{k-m+1})}},... \frac {s(1-\phi)-{\phi (1-\phi^{m-1})}}{{\phi^{m} (1-\phi^{k-m+1})}}}_{k-(m-1)},0...).
\end{align*}
Straightforward calculation shows that ${\bf z}^{0}_{m-1}$ is a convex combination of ${\bf z}^{0}_{m}$ and ${\bf z}^{1}_{m-1}$: ${\bf z}^{0}_{m-1}=\gamma^m {\bf z}^{0}_{m} + (1-\gamma^m) {\bf z}^{1}_{m-1}$, where
\begin{align*}
& {\bf z}^{0}_{m-1} = (\underbrace{\frac{sz_{m-1}}{s-\sum_{n=1}^{m-2}{\phi^n(z_n-z_{m-1})}},...\frac{sz_{m-1}}{s-\sum_{n=1}^{m-2}{\phi^n(z_n-z_{m-1})}}}_{m-1},\frac{sz_{m}}{s-\sum_{n=1}^{m-2}{\phi^n(z_n-z_{m-1})}},...), \\
& \gamma^m = \left(1-\frac{s z_{m-1}}{s-\sum_{n=1}^{m-2}\phi^n(z_n-z_{m-1})} \right) / \left( 1-\frac{s z_m}{s-\sum_{n=1}^{m-1}\phi^n(z_n-z_{m})} \right) \in [0,1].
\end{align*}

Similarly, we can write ${\bf z}^{0}_{j}$ as a convex combination of ${\bf z}^{0}_{j+1}$ and ${\bf z}^{1}_{j}$ for $j = 1,2,...m-1$: ${\bf z}^{0}_{j}=\gamma^{j+1} {\bf z}^{0}_{j+1} + (1-\gamma^{j+1}) {\bf z}^{1}_{j}$,  where
\begin{align*}
& {\bf z}^{0}_{j} = (\underbrace{\frac{sz_j}{s-\sum_{n=1}^{j-1}{\phi^n(z_n-z_j)}},...\frac{sz_j}{s-\sum_{n=1}^{j-1}{\phi^n(z_n-z_j)}}}_{j},\frac{sz_{j+1}}{s-\sum_{n=1}^{j-1}{\phi^n(z_n-z_j)}},...), \\
& {\bf z}^{1}_{j} = (\underbrace {1,...1}_{j}, \frac{(s(1-\phi)-{\phi (1-\phi^{j})}) z_{j+1}}{(s-\sum_{n=1}^{j}\phi^n z_n)(1-\phi)}, \frac{(s(1-\phi)-{\phi (1-\phi^{j})}) z_{j+2}}{(s-\sum_{n=1}^{j}\phi^n z_n)(1-\phi)},...), \\
& \gamma^{j+1} = \left(1-\frac{s z_{j}}{s-\sum_{n=1}^{j-1}\phi^n(z_n-z_{j})} \right) / \left( 1-\frac{s z_{j+1}}{s-\sum_{n=1}^{j}\phi^n(z_n-z_{j+1})} \right) \in [0,1].
\end{align*}
By deleting the first $j$ components of ${\bf z}^1_j$ we get a sub-vector ${\bf \hat z}^1_j \in \mathcal{Z}_- \left(\frac{s(1-\phi)-{\phi (1-\phi^{j})}}{(1-\phi)\phi^{j}}\right)$, which is a convex combination of ${\bf \hat z}^{j',k'}$, $0 \le j' \le m-j-1$ and $k' \ge m-j$, where
\begin{align*}
{\bf \hat z}^{j',k'} = (\underbrace {1,1,...1}_{j'}, \underbrace {\frac {s(1-\phi)-{\phi (1-\phi^{j+j'})}}{{\phi^{j+j'+1} (1-\phi^{k'-j'})}},... \frac {s(1-\phi)-{\phi (1-\phi^{j+j'})}}{{\phi^{j+j'+1} (1-\phi^{k'-j'})}}}_{k'-j'},0...).
\end{align*}
Equivalently, ${\bf z}^{1}_{j}$ is a convex combination of ${\bf z}^{l,k}$, $j \le l \le m-1$, $k \ge m$, where
\begin{align*}
{\bf z}^{l,k} = (\underbrace {1,1,...1}_{l}, \underbrace {\frac {s(1-\phi)-{\phi (1-\phi^{l})}}{{\phi^{l+1} (1-\phi^{k-l})}}, ...\frac {s(1-\phi)-{\phi (1-\phi^{l})}}{{\phi^{l+1} (1-\phi^{k-l})}}}_{k-l},0...).
\end{align*}

Finally, ${\bf z} = {\bf z}^{0}_{1}$ is a convex combination of ${\bf z}^{0}_2$ and ${\bf z}^{1}_{1}$. Then ${\bf z}^0_1$ is a convex combination of ${\bf z}^1_1, ..., {\bf z}^1_{m-1}$ and ${\bf z}^0_m$. Hence, ${\bf z}$  is a convex combination of ${\bf z}^{j,k}$, $0 \le j \le m-1$, $k \ge m$.
By a contradiction argument same as in the proof of Proposition \ref{prop:UnfairGames:infinites01}, the optimal solution ${\bf z}^* \in \cup_{j=0}^{m-1} \mathcal{Z}_-^j(s)$.
\halmos

\pfof{Proposition \ref{prop:UnfairGames:losspart:optimall1}}
For fixed $m$ and $j$, let
\begin{align}\label{eq:fmj}
f_{m,j}(x) = \left( (j+1+x)^{\alpha_-}-(j+1)^{\alpha_-} \right) \frac{1}{\left(1-\phi^x \right)^{\delta_-}}, \; x \geq 0.
\end{align}
Then
\begin{align*}
R_m(j,k) = \left((j+1)^\alpha - 1 \right) + \left( \frac {s{(1-\phi)}-{\phi (1-\phi^{j})}}{{\phi^{j+1}}} \right)^{\delta_-} f_{m,j}(k-j).
\end{align*}
Note that $f_{m,j}(x) \to +\infty$ as $x \to +\infty$ since $\phi < 1$.

The first-order derivative of $f_{m,j}$ is
\begin{align*}
f'_{m,j}(x) & = \alpha_- (j+1+x)^{\alpha_--1} \frac{1}{\left(1-\phi^x \right)^{\delta_-}} - \left( (j+1+x)^{\alpha_-}-(j+1)^{\alpha_-} \right) {\delta_-} \frac{\phi^x (-\log \phi)}{\left(1-\phi^x \right)^{\delta_- +1}} \\
& = \frac{\alpha_- (j+1+x)^{\alpha_- -1} x \phi^x (-\log \phi)}{\left(1-\phi^x \right)^{\delta_-+1}} \\
& \qquad \cdot \left(\frac{ 1-\phi^x }{x} \frac{1}{\phi^x (-\log \phi)} - \frac{\delta_-}{\alpha_-} \frac{ (j+1+x)^{\alpha_-}-(j+1)^{\alpha_-} }{x (j+1+x)^{\alpha_--1}} \right) \;.
\end{align*}

Proof of {\em (i)}.

On one hand, noting that $1-\phi^x$ is concave in $x \in (0,+\infty)$,
\begin{align*}
\frac{ 1-\phi^x }{x} \geq  \frac{d (1-\phi^x)}{d x} = {\phi^x (-\log \phi)}
\Rightarrow \frac{ 1-\phi^x }{x} \frac{1}{\phi^x (-\log \phi)} \ge 1.
\end{align*}
On the other hand,
\begin{align*}
(j+1+x)^{\alpha_-}-(j+1)^{\alpha_-} \leq (j+1+x)^{\alpha_-} - (j+1)(j+1+x)^{\alpha_- -1} =  x (j+1+x)^{\alpha_- -1}.
\end{align*}
Since $\delta_- \le \alpha_-$,
\begin{align*}
\frac{\delta_-}{\alpha_-} \frac{ (j+1+x)^{\alpha_-}-(j+1)^{\alpha_-} }{x (j+1+x)^{\alpha_- -1}} \leq 1.
\end{align*}
Hence, $f_{m,j}^{'}(x) \geq 0$.
As a result, $f_{m,j}(x)$ is increasing on $x$. For fixed $m$ and $j$, $R_m(j,k)$ is minimized at $k = m$.

Proof of {\em (ii)}.

Recall $f_{m,j}$ of \eqref{eq:fmj} and note that
\begin{align*}
f'_{m,j}(x) & = \frac{\alpha_- (j+1+x)^{\alpha_--1} \phi^x (-\log \phi)}{\left(1-\phi^x \right)^{\delta_- +1}} \left(\frac{ 1-\phi^x }{ \phi^x (-\log \phi)} - \frac{\delta_-}{\alpha_-} \frac{  (j+1+x)^{\alpha_-}-(j+1)^{\alpha_-} }{ (j+1+x)^{\alpha_- -1}} \right) \\
& = \frac{\alpha_- (j+1+x)^{\alpha_--1} \phi^x (-\log \phi)}{\left(1-\phi^x \right)^{\delta_- +1}} g_{m,j}(x) \;,
\end{align*}
where $g_{m,j}(x)$ is given in \eqref{eq:gmj}.
The first-order derivative of $g_{m,j}$ is
\begin{align*}
g'_{m,j}(x) = \frac{ 1 }{ \phi^x } - \frac{\delta_-}{\alpha_-} \left(1 - (j+1)^{\alpha_-} (1-\alpha_-)(j+1+x)^{-\alpha_-} \right) \;.
\end{align*}
The second-order derivative of $g_{m,j}$ is
\begin{align*}
g''_{m,j}(x) = \frac{ -\log \phi }{ \phi^x } - {\delta_-} (1-\alpha_-) (j+1)^{\alpha_-} (j+1+x)^{-\alpha_- -1} \;.
\end{align*}
Note that $g''_{m,j}(x)$ is increasing on $x$ and
\begin{align*}
g''_{m,j}(0) = -\log \phi - {\delta_-} (1-\alpha_-) (j+1)^{-1} \geq -\log \phi - {\delta_-} (1-\alpha_-).
\end{align*}
Then if $\phi \leq e^{-\delta_-(1-\alpha_-)}$, $-\log \phi - {\delta_-} (1-\alpha_-) \geq 0$ and hence, $g''_{m,j}(0) \ge 0$.
If $\phi > e^{-\delta_-(1-\alpha_-)}$ and $j \ge \delta_-(1-\alpha_-)/(-\log\phi) -1$, then we also have $g''_{m,j}(0) \ge 0$.
As $g''_{m,j}(x)$ is increasing on $x$, $g''_{m,j}(x) \ge g''_{m,j}(0) \ge 0$.
As a result, $g'_{m,j}(x)$ is increasing on $x$.
Note that $g'_{m,j}(0) = 1-\delta_- \geq 0$ and $g'_{m,j}(x) \to +\infty$ as $x \to +\infty$.
Then $g'_{m,j}(x) \geq 0$ for $x \in (0,+\infty)$. Hence, $g_{m,j}(x)$ is increasing on $x$. Combing with the fact that $g_{m,j}(0) = 0$, we have $f'_{m,j}(x) \geq 0$ for $x \in (0,+\infty)$. Consequently, $f_{m,j}(x)$ is increasing on $x$. Therefore, $R_m(j,k)$ is minimized at $k = m$.

If $\phi > e^{-\delta_-(1-\alpha_-)}$ and $j < \delta_-(1-\alpha_-)/(-\log\phi) -1$, $g''_{m,j}(0) < 0$. As $g''_{m,j}(x)$ is strictly increasing on $x$ and $g''_{m,j}(x) \to +\infty$ as $x \to +\infty$, there exists $x_2 > 0$ such that $g''_{m,j}(x) <  0$ for $x < x_2$ and $g''_{m,j}(x) > 0$ for $x > x_2$. Hence, $g'_{m,j}(x)$ is strictly decreasing on $x < x_2$ and strictly increasing on $x > x_2$.

Suppose $\delta_- < 1$. If $g'_{m,j}(x_2) \ge 0$, then $g'_{m,j}(x) \ge 0$ for all $x \ge 0$ and therefore, $g_{m,j}(x)$ is increasing on $x$. As a result, with the same reason as above,  $f_{m,j}(x)$ is increasing on $x$ and $R_m(j,k)$ is minimized at $k = m$.
If $g'_{m,j}(x_2) < 0$, then there exist $x_{1l}$ and $x_{1r}$ such that $g'_{m,j}(x_{1l}) = g'_{m,j}(x_{1r}) = 0$. Therefore, $g'_{m,j}(x) > 0$ for $x < x_{1l}$ and $x > x_{1r}$, $g'_{m,j}(x) < 0$ for $x \in (x_{1l},x_{1r})$.
As a result, $g_{m,j}(x)$ is strictly increasing on $x < x_{1l}$, then strictly decreasing on $x_{1l} < x < x_{1r}$, and then strictly increasing on $x > x_{1r}$. If $g_{m,j}(x_{1r}) \ge 0$, then $g_{m,j}(x) \ge 0$ for all $x \ge 0$. With the same reason as above,  $f_{m,j}(x)$ is increasing on $x$ and $R_m(j,k)$ is minimized at $k = m$.

If $g_{m,j}(x_{1r}) < 0$, then there exists $x_{0l}$ and $x_{0r}$ such that $g_{m,j}(x_{0l}) = g_{m,j}(x_{0r}) = 0$, $g_{m,j}(x) > 0$ for $x < x_{0l}$ and $x > x_{0r}$, $g_{m,j}(x) < 0$ for $x \in (x_{0l},x_{0r})$.
Therefore, $f_{m,j}(x)$ is strictly increasing on $x < x_{0l}$, then strictly decreasing on $x_{0l} < x < x_{0r}$, and then strictly increasing on $x > x_{0r}$. Let $x_0 = x_{0r}$. If $j + \lfloor x_0 \rfloor \ge m+1$,  $R_m(j,k)$ is minimized at $k \in \{m, j+\lfloor x_0 \rfloor, j + \lfloor x_0 \rfloor + 1\}$. If $j + \lfloor x_0 \rfloor = m$,  $R_m(j,k)$ is minimized at $k \in \{m, m + 1\}$. If $j + \lfloor x_0 \rfloor \le m-1$,  $R_m(j,k)$ is minimized at $k = m$.

On the other hand, suppose $\delta_- = 1$. Then there exist $x_{1} > 0$ such that $g'_{m,j}(x_{1}) = 0$. Therefore, $g'_{m,j}(x) < 0$ for $x \in (0,x_{1})$ and $g'_{m,j}(x) > 0$ for $x > x_{1}$. As a result, $g_{m,j}(x)$ is strictly decreasing on $0 < x < x_{1}$, and then strictly increasing on $x > x_{1}$. Then there exists $x_{0} > 0$ such that $g_{m,j}(x_{0}) = 0$, $g_{m,j}(x) < 0$ for $x \in (0,x_{0})$, and $g_{m,j}(x) > 0$ for $x > x_{0}$. Therefore, $f_{m,j}(x)$ is strictly decreasing on $0 < x < x_{0}$, and then strictly increasing on $x > x_{0}$.  If $j + \lfloor x_0 \rfloor \ge m+1$,  $R_m(j,k)$ is minimized at $k \in \{m, j+\lfloor x_0 \rfloor, j + \lfloor x_0 \rfloor + 1\}$. If $j + \lfloor x_0 \rfloor = m$,  $R_m(j,k)$ is minimized at $k \in \{m, m + 1\}$. If $j + \lfloor x_0 \rfloor \le m-1$,  $R_m(j,k)$ is minimized at $k = m$.
\halmos


\pfof{Proposition \ref{prop:mergedproblem}}
To solve \eqref{prob:UnfairGames:OptimalProbabilityDistribution}, first fix $s_+$ and $s_-$ and write $z_{+,1}$ as $\phi\frac{s_-+1}{s_++1}z_{-,1}$.
Then the objective of \eqref{prob:UnfairGames:OptimalProbabilityDistribution} is written as $f(z_{-,1})$, with constraint $0 \leq z_{-,1} \leq \frac{s_++1}{s_++1+\phi(s_-+1)}$.


The first-order derivative of $f$ is
\begin{align*}
f'(y) 
= y^{\delta_--1} \left({\delta_+}(v_{+}^*(s_+)+1) \phi^{\delta_+} \left(\frac{s_-+1}{s_++1} \right)^{\delta_+} y^{\delta_+-\delta_-} - \lambda {\delta_-} (v_{-}^*(s_-)+1) \right).
\end{align*}

{\it (i)} Suppose $\delta_+ \geq \delta_-$. Then $f'(y) / y^{\delta_--1}$ is increasing on $y$. Note that $\lim_{y \searrow 0} f'(y) < 0$. As a result, $f(\cdot)$ is either decreasing, or first decreasing and then increasing on $y$. Then $f(y)$ is maximized when $y = 0$ or $y = \frac{s_++1}{s_++1+\phi(s_-+1)}$. Note that
\begin{align*}
& \quad f \left(\frac{s_++1}{s_++1+\phi(s_-+1)} \right) \\
& =  (v_{+}^*(s_+)+1) \phi^{\delta_+}\left(\frac{s_-+1}{s_++1+\phi(s_-+1)} \right)^{\delta_+}
-\lambda (v_{-}^*(s_-)+1)\left({\frac{s_++1}{s_++1+\phi(s_-+1)}} \right)^{\delta_-} \\
& = (v_{-}^*(s_-)+1)\left({\frac{s_++1}{s_++1+\phi(s_-+1)}} \right)^{\delta_-} \left(\frac{(v_{+}^*(s_+)+1) \phi^{\delta_+} \left({s_-+1} \right)^{\delta_+}}{(v_-^*(s_-)+1)\left({{s_++1}} \right)^{\delta_-} (s_++1+\phi(s_-+1))^{\delta_+-\delta-}} - \lambda \right) \\
& \le (v_{-}^*(s_-)+1)\left({\frac{s_++1}{s_++1+\phi(s_-+1)}} \right)^{\delta_-} (M - \lambda).
\end{align*}

If $\lambda \geq M$, then $f(\frac{s_++1}{s_++1+\phi(s_-+1)}) \leq 0$. Hence, $f(y)$ is maximized when $y = 0$. In this case, $z_{+,1}^* = z_{-,1}^* = 0$, meaning that the gambler does not play the game.

On the other hand, if $\lambda < M$, the optimal value of $f(y)$ is positive, which means the gambler will play the game. $f(y)$ is maximized when $y = y^*(s_+,s_-) = \frac{s_++1}{s_++1+\phi(s_-+1)}$. With $s_+^*$ and $s_-^*$ maximizing $f(\frac{s_++1}{s_++1+\phi(s_-+1)})$, the optimal solution to (\ref{prob:UnfairGames:OptimalProbabilityDistribution}) is $z_{-,1}^* =  y^*(s_+^*,s_-^*)$ and $z_{+,1}^* = \phi\frac{s_-^*+1}{s_+^*+1} z_{-,1}^*$, which is given by \eqref{eq:sol1}.

{\it (ii)} Suppose $\delta_+ < \delta_-$. Then $f'(y)/ y^{\delta_--1}$ is decreasing on $y$. Note that $\lim_{y \searrow 0} f'(y) > 0$. As a result, $f(\cdot)$ is either increasing, or first increasing and then decreasing on $y$.
Note that $f'(y)/ y^{\delta_--1} = 0$ when
\begin{align*}
y = \bar y(s_+,s_-) := \left( \frac{{\delta_+} (v_{+}^*(s_+)+1) \phi^{\delta_+}\left({s_-+1} \right)^{\delta_+}}{\lambda {\delta_-} (v_{-}^*(s_-)+1) \left({s_++1} \right)^{\delta_+} } \right)^{\frac{1}{\delta_--\delta_+}} \;.
\end{align*}
Then $f(y)$ is maximized when
\begin{align*}
y = k(s_+,s_-) = \min \left\{\bar y(s_+,s_-), \frac{s_++1}{s_++1+\phi(s_-+1)} \right\} \;.
\end{align*}
In this case the optimal value of $f(y)$ is positive, meaning that the gambler will play the game. With $s_+^*$ and $s_-^*$ maximizing $f(k(s_+,s_-))$, the optimal solution to (\ref{prob:UnfairGames:OptimalProbabilityDistribution}) is $z_{-,1}^* = k(s_+^*,s_-^*)$ and $z_{+,1}^* = \phi\frac{s_-^*+1}{s_+^*+1} z_{-,1}^*$, which is given by \eqref{eq:sol2}.
\halmos

\end{document}